\documentclass[%
 aip,
 jcp,
 amsmath,amssymb,
 reprint,%
]{revtex4-1}

\usepackage{graphicx}
\usepackage{dcolumn}
\usepackage{bm}

\usepackage[utf8]{inputenc}
\usepackage[T1]{fontenc}
\usepackage{mathptmx}
\usepackage{etoolbox}
\usepackage{multirow}
\usepackage[caption=false,justification=centering]{subfig}
\usepackage[colorlinks=true,linkcolor = blue,urlcolor=blue,citecolor=blue]{hyperref}

\makeatletter
\def\@email#1#2{%
 \endgroup
 \patchcmd{\titleblock@produce}
  {\frontmatter@RRAPformat}
  {\frontmatter@RRAPformat{\produce@RRAP{*#1\href{mailto:#2}{#2}}}\frontmatter@RRAPformat}
  {}{}
}%
\makeatother
\begin{document}

\preprint{AIP/123-QED}
\title{Machine learning for accelerated bandgap prediction \\in strain-engineered quaternary III-V semiconductors} 
\author{Badal Mondal}
\affiliation{Wilhelm-Ostwald-Institut f\"ur Physikalische und Theoretische Chemie, Universit\"at Leipzig, 04103 Leipzig, Germany}
\affiliation{Fachbereich Physik, Philipps-Universit{\"a}t Marburg, 35032 Marburg, Germany}

\author{Julia Westermayr}
\affiliation{Wilhelm-Ostwald-Institut f\"ur Physikalische und Theoretische Chemie, Universit\"at Leipzig, 04103 Leipzig, Germany}
\affiliation{Center for Scalable Data Analytics and Artificial Intelligence, Dresden/Leipzig, Germany}

\author{Ralf Tonner-Zech}
\affiliation{Wilhelm-Ostwald-Institut f\"ur Physikalische und Theoretische Chemie, Universit\"at Leipzig, 04103 Leipzig, Germany}
\email{ralf.tonner@uni-leipzig.de}

\date{\today}
\begin{abstract}
Quaternary III-V semiconductors are one of the major promising material classes in optoelectronics. The bandgap and its character, direct or indirect, are the most important fundamental properties determining the performance and characteristics of optoelectronic devices. Experimental approaches screening a large range of possible combinations of III- and V-elements with variations in composition and strain are impractical for every target application. We present a combination of accurate first-principles calculations and machine learning based approaches to predict the properties of the bandgap for quaternary III-V semiconductors. By learning bandgap magnitudes and their nature at density functional theory accuracy based solely on the composition and strain features of the materials as an input, we develop a computationally efficient yet highly accurate machine learning approach that can be applied to a large number of compositions and strain values. This allows for a computationally efficient prediction of a vast range of materials under different strains, offering the possibility for virtual screening of multinary III-V materials for optoelectronic applications.
\end{abstract}

\maketitle


\section{\label{sec:introduction} Introduction}
Semiconductor compounds are central to modern optoelectronics and find applications in various fields, such as solar cells, light-emitting diodes, optical telecommunication, and photovoltaics.\cite{Soref1993Silicon,CardanoBook,PhilippsS2018Chapter,Geisz2002,Mokkapati2009IIIV,Dimroth2016Four,Sato2002,Fuchs2018,Hepp2022Room,Sweeney2011,Wang2013} One of the fundamental properties determining the performance of such optoelectronic devices is the bandgap. The tuning of the size and type of bandgaps is one of the major goals in the field of optoelectronics. Varying the relative composition in compound semiconductors is one of the major approaches here.\cite{Beyer2015Metastable,Vurgaftman2001Band,Beyer2017Local,Ludewig2016Movpe,Liebich2011Laser,Supplie2018Metalorganic,Stringfellow2019Fundamental,Volz2009Movpe,Feifel2017Movpe,Kunert2004Movpe,Volz2004Specific,Veletas2019Bismuth,Wegele2016Interface,Hepp2019Movpe} Alternatively, straining the system can be used to modify the bandgaps.\cite{Van1990Pressure,Potter1956Indirect,Alekseev2020Effect,Katiyar2020Breaking,Lim2021Strain,Signorello2014Inducing,Signorello2013Tuning,Balaghi2019Widely,Gronqvist2009Strain,Hetzl2016,Montazeri2010,Skold2005} By combining these two approaches, bandgaps can be tailored over a wide range of values, enabling the enormous diversity in device applications.\cite{Beyer2015Metastable,Vurgaftman2001Band,Beyer2017Local,Ludewig2016Movpe,Liebich2011Laser,Supplie2018Metalorganic,Stringfellow2019Fundamental,Volz2009Movpe,Feifel2017Movpe,Kunert2004Movpe,Volz2004Specific,Veletas2019Bismuth,Wegele2016Interface,Hepp2019Movpe,Bir1974Symmetry,Sun2007Physics,Tao2020,Tsutsui2019,Fang2011} Due to a vast composition space, quaternary III-V semiconductors offer a unique opportunity in materials design.\cite{Guden1996} However, identifying tailored materials for each target application requires assessing the dependence of the bandgap on composition and strain for a large set of materials. Because of the tremendous effort necessary for the synthesis of unknown materials, experimental approaches of screening the vast chemical space of all possible combinations of III-(or group 13) and V-(or group 15) elements with variation in composition and strain, thus, are not practical.\cite{Beyer2015Metastable,Liebich2011Laser,Volz2009Movpe,Volz2004Specific,Wegele2016Interface,Hepp2019Movpe} Therefore, theoretical models that are both accurate and computationally efficient are often the only viable choice for high-throughput virtual screening for materials design.\cite{Westermayr2023,Sanchez-Lengeling2018}

In recent years, density functional theory (DFT) methods based on computationally efficient density functionals like those based on the local density approximation (LDA)\cite{Kohn1965} or generalized gradient approximation (GGA)\cite{Perdew1996Generalized} have proven to be powerful and successful tools for such high-throughput material screening. However, large errors for semiconductor bandgaps, which can be in the range of 50\% of the bandgap value,\cite{Mori2008,Rosenow2018Ab,Koller2011Merits} are common. Better accuracy can be achieved with methods such as hybrid functionals,\cite{Krukau2006Influence,Heyd2003} many-body perturbation theory (GW),\cite{Hybertsen1986Electron,Hedin1965New,Aryasetiawan1998Gw} or meta-GGA functionals such as the modified Becke-Johnson functional (TB09).\cite{Tran2009Accurate} Our previous study on mapping bandgaps for binary and ternary III-V semiconductors using the TB09 functional over a large range of composition and strain values showed excellent agreement with experiment at moderate computational costs.\cite{Mondal2022,Mondal2023} However, due to the large composition space of quaternary III-V semiconductor materials leading to an estimated amount of 10,000,000s\cite{number_dft_estimate} of calculations per combination elements, the computational costs of these methods are still too large to enable high-throughput studies. 

To speed up the exploration of chemical space, machine learning (ML) techniques have been applied in various fields of material science, such as the prediction and classification of crystal structures,\cite{Butler2018,Fischer2006,Carr2009,Pilania2015,Yamashita2018,Gu2006,Rajan2018} thermal properties,\cite{Seko2014,Pilania2015,Gu2006,Ward2016} electronic properties,\cite{Rajan2018,Schutt2014,Li2022,Wang2022,Prateek2023,Ward2016,Pilania2016,Heng2000,Gladkikh2020,Zhuo2018,Gu2006,Lee2016,Wu2020,Wang2021,Weston2018,Venkatraman2021,Westermayr2021,Ghosh2019} and stability of materials.\cite{Ye2018,Schmidt2017,Ward2016,Gu2006,Heng2000} Among the most widely used ML models in the field of bandgap predictions of semiconductor materials are support vector machine (SVM) models.\cite{Lee2016,Huang2019,Zhu2020,Weston2018,Zhuo2018,Gu2006} However, for III-V semiconductors, the previous reports using ML methods predicted the bandgap values of unstrained compounds only.\cite{Zhu2020,Huang2019,Zhuo2018,Lee2016,Heng2000,Gu2006} No ML study on strained systems is available so far. Moreover, due to the large compositional space, the bandgap values of only a few selected III-V ternary and quaternary compositions have been predicted so far with ML methods.\cite{Huang2019,Heng2000,Zhuo2018}

In this work, we go beyond previous approaches and develop an ML model for bandgap predictions in biaxially strained quaternary III-V materials (AB$_x$C$_y$D$_{100-x-y}$; with A = III element; B, C, D = V elements) over the complete composition range ($x,y = 0-100$) and a compressive and tensile strain range of $5\%$ around the unstrained structures. The excellent performance of the approach is demonstrated. Subsequently, we construct the composition-strain-bandgap relationship, the ``bandgap phase diagram'',\cite{Mondal2023,bandgap_phase_diagram_github} for GaAsPSb using our ML model. Although many of the binary and ternary subsystems of GaAsPSb, namely GaAs, GaP, GaSb, GaAsP, GaAsSb, and GaPSb, have been successfully synthesized and found special applications in different research fields,\cite{Craford1973,Henning1983,Tanaka1994,Sato2002,Geisz2002,Lang2013,Hayashi1994,Grassman2016,Rosenow2018Ab,Weyers1992,Kunert2006,Zhao2004,Loualiche1998,Shimomura1996,Nakajima2000,Russell2016,Jou1988,Cherng1984,Jen1998} this particular quaternary compound has not been studied yet. Therefore, our theoretical predictions can provide insights for future experimental exploration of this material system. 

We emphasize that the predictive capability of the ML models relies on the quality of the reference method used in the generation of ML training data. Given the lack of sufficient experimental data, we opted to use DFT-based methods to create the training dataset for our ML model in this instance. Since our DFT computational setup, as used here, has consistently demonstrated high accuracy in generating bandgaps that align well with experimental data,\cite{Rosenow2018Ab,Koller2011Merits,Tran2009Accurate,Mondal2022,Mondal2023} our ML model serves as a valuable tool for the initial screening of the vast composition-strain space to guide the next computational and experimental steps. Additionally, our approach allows for straightforward extension to other III-V quaternary systems and, thus, provides a general theoretical basis for targeted exploration of this compound class. The integrated first-principles calculations and ML techniques provide a powerful approach for future computational materials design. The ultimate goal of this work is to provide comprehensive guidelines for studying bandgap properties in strained materials.

The article is organized as follows. In Sec.~\ref{sec:computationaldetails}, we discuss the methodology: details of DFT calculations (Sec.~\ref{subsec:dftdetails}) and an overview of ML methods (Sec.~\ref{subsec:mldetails}). Next, we present the results in Sec.~\ref{sec:results} and establish the best ML models and hyperparameters for predicting the bandgaps of strained GaAsSbP. Using these methods, a complete bandgap phase diagram is constructed for GaAsSbP in Sec.~\ref{subsec:gaaspsbbpd}. Finally, we summarize our key findings in Sec.~\ref{sec:summary}.

\section{\label{sec:computationaldetails}Methods}
\subsection{\label{subsec:dftdetails}First-principle computational details}
First, we created a dataset of 4280 data points obtained from DFT encompassing the whole composition and strain range of GaAsPSb investigated. The dataset comprises 88 calculations of the corresponding strained binary subsystems of GaAsSbP, namely, GaAs, GaP, and GaSb in total (taken from Ref.~\citenum{Mondal2022}); 2272 composition-strain data points corresponding to the ternary subsystems of GaAsPSb, namely, GaAsP, GaPSb, and GaAsSb (taken from Ref.~\citenum{Mondal2023}); and 1920 randomly chosen points in the composition-strain space of GaAsPSb, containing non-zero percentage of all elements As, P, and Sb. Notably, only the data for biaxially strained structures are collected from the Refs.~\citenum{Mondal2022} and \citenum{Mondal2023}. More details on the data features are discussed in the next section. The complete dataset can be found in the Supplementary Material (SM) attachment. For convenience, in the following, we label the above three datasets as binary, ternary, and quaternary datasets, respectively.

The DFT computational setup for the quaternary calculations follows the approach used in the binary and ternary datasets calculations in our previous studies.\cite{Mondal2022,Mondal2023} The computations are performed using the projector-augmented wave (PAW) method~\cite{Kresse1999From,Blochl1994Projector} as implemented in the Vienna \textit{ab-initio} simulation package (VASP). \cite{Kresse1993Ab, Kresse1994Ab, Kresse1996Efficient, Kresse1996Efficiency} The generalized gradient approximation (GGA) based exchange-correlation functional by Perdew, Burke, and Ernzerhof (PBE)~\cite{Perdew1996Generalized} with a cut-off energy of 550 eV for the planewave basis set is chosen in all calculations. Corrections for missing dispersion interactions are calculated using the semiempirical DFT-D3 approach with improved damping function.\cite{Grimme2010Consistent,Grimme2011Effect} The electronic energy convergence criteria of $10^{-7}$ eV and the force convergence of $10^{-2}$ eV\AA$^{-1}$ are used, respectively. The quaternary materials are modeled using special quasirandom structures (SQS)~\cite{Zunger1990Special} with the supercell of size $6\times 6\times 6$. The SQS cells are generated using the alloy-theoretic automatic toolkit (ATAT).\cite{VandeWalle2002, VandeWalle2009, VanDeWalle2013} The reciprocal space is sampled at the $\Gamma$-point only, which is sufficient due to the large real-space cell size. Geometry optimizations are carried out by consecutive volume and position optimization until convergence is reached. For the bandgap calculations the meta-GGA functional TB09 \cite{Tran2009Accurate} is used, including spin-orbit coupling. The bandgap natures are determined using the Bloch-spectral-weight-based protocol as described in Ref.~\citenum{Mondal2023}. The `fold2Bloch'~\cite{Rubel2014} code is used to determine the Bloch spectral weights. 

The materials investigated here all feature the zincblende-type structures only. Moreover, since these compounds can only be experimentally realized through epitaxy, we only consider the effect of biaxial strain. As [100] crystal direction is the most common choice of substrate orientation and the crystal growth direction in epitaxy, we model the strain application along [100] directions. The in-plane lattice parameters ([100] and [010] directions) are kept fixed at the chosen strain values while the out-of-plane direction [001] is relaxed. The strained in-plane lattice parameters ($a_f$) are calculated using the following formula (Eq.~\ref{eqn:eqn1}):
\begin{eqnarray}
a_f = a_{eqm} \times \left(1 + \frac{\textrm{biaxial strain (\%)}}{100} \right) \label{eqn:eqn1}
\end{eqnarray}
where $a_{eqm}$ is the equilibrium lattice parameter of the specific structure that will be strained.\cite{Mondal2022,Mondal2023} Following the convention from Refs.~\citenum{Mondal2022} and \citenum{Mondal2023}, in the following, we indicate tensile strain with positive sign and compressive strain with negative sign. No structural phase transition is assumed under strain application. Notably, we limit our analyses within compressive and tensile strain range of $5\%$. This is typically the strain range achievable via epitaxial growth. Moreover, our analysis assumes perfect epitaxial growth and does not account for defects.

\subsection{\label{subsec:mldetails}Machine learning details}
\subsubsection{\label{subsec:mlmodels}ML Model choice}
We use the support vector regression (SVR)~\cite{Smola2004} model to train and predict bandgap values. For learning and predicting the bandgap nature, we use the support vector classification (SVC) model.\cite{Cortes1995,Burges1998} Below we present the key features of each of these SVM models. All the ML algorithms are used as implemented in \verb|scikit-learn|.\cite{JMLR:v12:pedregosa11a:short}

The fundamental feature of SVM is that it constructs a (set of) hyper-plane(s) in a high dimensional space. The main objective here is to optimize the positions of these hyper-planes to effectively separate the training data points based on their classes. A good separation is defined when the hyper-plane has the largest distance to the nearest training data points of any class, the so-called functional margin, since maximizing the margin lowers the model's generalization error. However, achieving a perfect separation is not always feasible in real-world datasets. Therefore, an additional regularization parameter, $C$, is introduced to construct the so-called `soft margin'. 

Moreover, a nonlinear transformation is applied to the feature space by the so-called ``kernel trick''.\cite{scholkopf2018learning,Hofmann2008,Vempati2010} The kernel trick simplified the learning task by efficiently mapping the original feature space of the considered data into the new space using a kernel function, thus significantly reducing the cost of learning with large datasets. Here, we use the radial basis function (RBF) kernel,\cite{scholkopf2018learning,Vempati2010} which is defined as follows:
\begin{equation}
    k\left(\mathbf{x}_i, \mathbf{x}_j\right) = e^{-\gamma || \mathbf{x}_i- \mathbf{x}_j ||^2} \label{eqn:eqn7}
\end{equation}
The parameter $\gamma$ determines the inverse of the area of influence of the $i$-th data and decays with the distance to another $j$-th data. A low value of $\gamma$ means the influence reached `far', and a high value means `close'. The behavior of the model is very sensitive to the $\gamma$ values and can be tuned to optimize predictions.  

This defines a constrained optimization problem for SVM classification (SVC):
\begin{align}
&{\underset {\omega ,\;b,\;\mathbf {\xi } }{\textrm{minimize}} } && \frac{1}{2}||\omega||^2 + C\sum_{i=1}^{N} \xi_i\label{eqn:eqn2}\\ 
&\textrm{subject to} && y_i(\langle\omega , \phi(\mathbf{x}_i)\rangle + b) \geq 1-\xi_i \nonumber \\
&&& \xi_i\geq 0,\quad i=1,\ldots,N \label{eqn:eqn3}
\end{align}
where $\mathbf{x}_i$ is the feature vector (see Sec.~\ref{subsubsec:featurespace}) and $y_i$ refers to the original value of the $i$-th data point from a dataset $\{(\mathbf{x}_i;y_i)\}_{i=1}^N$. In our context, $y_i$ takes on the values of either +1 or --1, indicating whether the bandgaps obtained from quantum chemistry calculations are direct or indirect. The nonlinear mapping function $\phi$ maps the features $\mathbf{x}_i$ to the higher-dimensional space, which can be efficiently performed using kernel trick. The misclassification error for the $i$-th data point during optimization is measured by $\xi_i$. Its value is zero if the data point is correctly classified. Otherwise, the data points on the wrong side get a penalty. The $\omega$ and $b$ define the weight vector and offset values, respectively, which are optimized as indicated in Eq.~\ref{eqn:eqn2}. The minimization of $\frac{1}{2}||\omega|| ^2$ maximizes the margin. The $C$ parameter, thus, trades off the misclassification penalty of training samples against the maximization of the margin. 

Similarly, in the case of regression (SVR), the optimization problem is formulated as follows: 
\begin{align}
&{\underset {\omega ,\;b,\;\xi,\;\xi^* }{\textrm{minimize}} } &&\frac{1}{2}||\omega||^2 + C\sum_{i=1}^{N} (\xi_i+\xi^*_i)\label{eqn:eqn4}\\ 
&\textrm{subject to} &&y_i - (\langle\omega , \phi(\mathbf{x}_i)\rangle + b) \leq \varepsilon+\xi_i \nonumber\\
&&&(\langle\omega , \phi(\mathbf{x}_i)\rangle + b) - y_i \leq \varepsilon+\xi^*_i \nonumber \\
&&&\xi_i, \;\xi^*_i \geq 0\label{eqn:eqn5}
\end{align}
Here, an additional parameter, $\varepsilon$, is introduced to define the margin size around the predicted regression function. It determines the tolerance for errors in the regression model. Data points for which predictions ($\langle\omega , \phi(\mathbf{x}_i)\rangle + b$) are within the $\varepsilon$-tube are considered to have acceptable errors and do not contribute to the penalty term in the objective function. Otherwise, a penalty is added, determined by $\xi$ or $\xi^*$, depending on whether their predictions lie above or below the $\varepsilon$-tube. 

Once the optimization problem is solved, the prediction ($\hat{y}$) for a (new) sample \textbf{z} can be performed using the optimized $\omega$ and $b$:
\begin{equation}
    \hat{y} = \langle\omega , \phi(\mathbf{z})\rangle + b\label{eqn:eqn6}
\end{equation}
For classification, the predicted class corresponds to the sign of prediction ($\operatorname{sign}(\hat{y})$).

\subsubsection{\label{subsubsec:hyperparameteroptimization}Hyperparameter optimization}
A proper choice of the hyperparameters is critical to the SVM model's performance and accuracy. We thus train multiple ML models with 5-fold cross-validation at several random combinations of the hyperparameters ($C$ and $\gamma$ for SVC-RBF; $C$, $\gamma$, and $\varepsilon$ for SVR-RBF) values. The model exhibiting the highest cross-validation score is then picked as the final model holding the best set of hyperparameters. We utilize the \verb|RandomizedSearchCV| from \verb|scikit-learn| to automatize this hyperparameter optimization routine.

It is important to note that the specific combination of optimized hyperparameter values ($C$, $\gamma$, and $\varepsilon$ combination) may vary upon rerunning the hyperparameter optimization routine due to the stochastic nature of the random search and the random splitting of training and validation datasets within the cross-validation routine. We thus start with a hyperparameter optimization space expanding a large $C$, $\gamma$, and $\varepsilon$ intervals and carefully tune the ranges until the optimal ranges are found, covering the regimes with the best hyperparameters set. This ensures consistent performance of final best models (i.e, cross-validation scores do not change significantly) despite changes in the exact combination of optimized hyperparameter values upon rerunning the optimization routine.

Figure~S1 in the Supporting Information (SI) shows an example of the tuning of hyperparameter space for the SVR-RBF model. We perform a random search over uniformly distributed (in log-scale) 1000 random combinations of the $C$, $\gamma$, and $\varepsilon$ values. In this case, the optimal ranges found are 0.01--1000, 0.01--10, and 0.01--1 for the $C$, $\gamma$, and $\varepsilon$ ranges, respectively (SI Fig.~S1d).

\subsubsection{\label{subsubsec:modelperfrmancemetric}Performance metric}
The prediction accuracies of the SVR-RBF learning models are evaluated by means of root-mean-squared error (RMSE), mean absolute error (MAE), maximum error (Max error), and the coefficient of determination (R$^2$) metric. The SVC-RBF models are tested using accuracy-score and balanced-accuracy-score performance metrics. The hyperparameters are optimized using the RMSE and accuracy-score scoring functions for SVR-RBF and SVC-RBF models, respectively. We find some of the SVR-RBF models predict small negative direct bandgap values, up to --5 meV (see left-bottom corner in SI Fig.~S2). In this manuscript, we do not consider possible physical effects leading to negative bandgaps, such as topological band inversion, in our DFT dataset. We thus shift all the ML predicted negative bandgap values to 0 eV (SI Fig.~S2).

\subsubsection{\label{subsubsec:featurespace}Feature space and data preprocessing}

\begin{figure*}[!htbp]
\centering
 \subfloat[]{\label{fig:fig1a}\includegraphics[height=6.2cm]{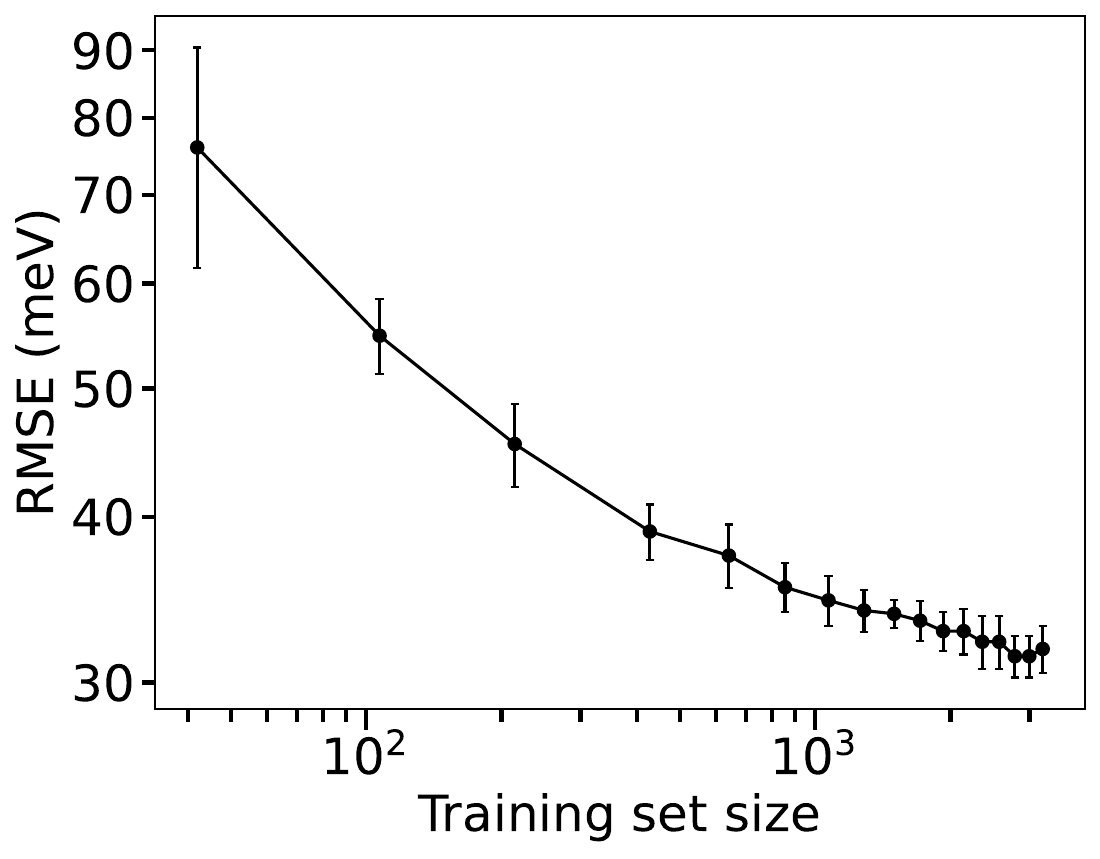}}
 \hspace{1.cm}
 \subfloat[]{\label{fig:fig1b}\includegraphics[height=6.2cm]{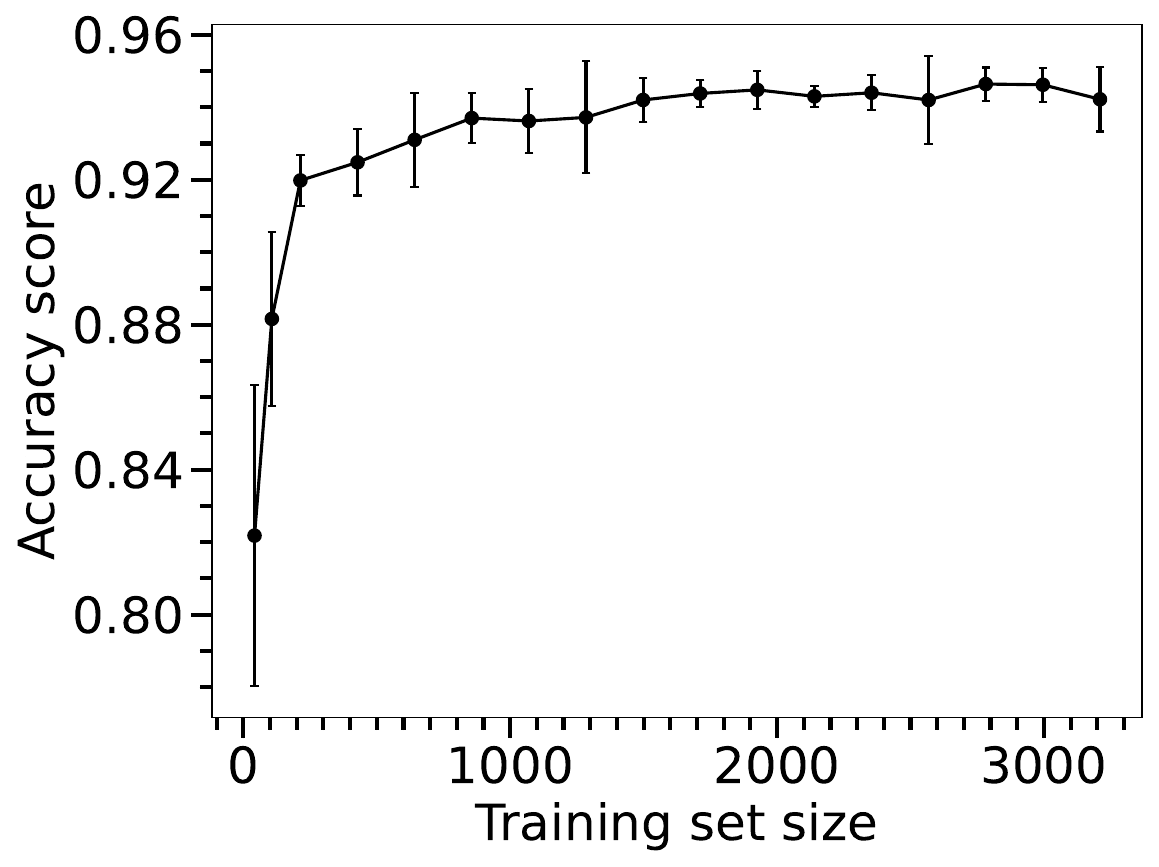}}
   \caption{Dependence of SVM-RBF models out-of-sample performances on the training set size. (a) Bandgap magnitude prediction RMSE from SVR-RBF model, in log--log scale. The hyperparameters are optimized with RMSE metric. (b) Bandgap nature prediction accuracy from SVC-RBF model. The hyperparameters are optimized with accuracy-score metric. Error bars show standard deviations over 5 trials.}
   \label{fig:fig1}
\end{figure*}
Alongside the choice of an appropriate model, another crucial part in ML is the choice of descriptors to represent the system under investigation. A number of different feature representations have been proposed for periodic solid-state systems, such as element-specific features,\cite{Schmidt2017,Gu2006,Zhuo2018,Pilania2016,Ward2016,Zhu2020,Gladkikh2020} radial distribution functions,\cite{Schutt2014} Voronoi tessellations,\cite{Ward2017} representation learning from stoichiometry,\cite{Goodall2020} and property-labeled materials fragments.\cite{Isayev2017} As our analysis consists of the prediction for strained and unstrained structures, we resort to keeping the descriptor as simple as possible and have chosen the composition and strain values as input features for the ML training. This allows for fast prediction and training times compared to the use of extensive descriptors and is found to be accurate enough for this study. The composition features can generally be constructed considering all the III- and V-elements. However, in the present article, we only present the results for GaAsPSb. In this case, the number of compositional degrees of freedom is reduced to only 3 ($x$, $y$, and $z$ in GaAs$_x$P$_y$Sb$_z$). Although only $x$ and $y$ are independent compositional degrees of freedom here, and $z$ can be deduced from $x$ and $y$ ($z=100-x-y$), adding $z$ in the feature vector ensures the model learns this constraint. Together with the measure of biaxial strain, the final feature space thus is only 4-dimensional (see SI Sec.~S3). As shown in the next section, this 4-dimensional feature space performs exceptionally well. We note that as a drawback, the above ML model is limited to the elements and type of strain (biaxial strain) chosen. Still, in principle, the descriptor could simply be extended to cover all possible chemical elements and strain types, allowing for universal application of the model.

Moreover, the features in our dataset, namely composition and strain, have different orders of magnitude in variance. Therefore, before training the models, the input data are standardized using the \verb|StandardScaler| class from \verb|scikit-learn|. 
 
\subsubsection{\label{subsubsec:datasetconvergence_modellingsec}Dataset size convergence}
We construct the learning curve to ensure the proper training of models and the comprehensiveness of the training set. Therefore, we create a test set consisting of random 25\% of the total input dataset comprising 4280 data points. We create the training set from the rest 75\% of the data, consecutively increasing the set size from 1\% up to 75\%. The series is repeated 5 times. We use 5-fold \verb|ShuffleSplit| from \verb|scikit-learn| to perform the above train-test splittings. For each ML model training, hyperparameters are re-optimized.


\section{\label{sec:results}Results and discussion}
Figure~\ref{fig:fig1} shows the dependence of SVM-RBF model out-of-sample performances on the training set size. As the training set size increases, the RMSE of the SVR-RBF model decreases by approximately 45 meV(Fig.~\ref{fig:fig1a}). Eventually, it reaches $\sim$30 meV with the largest training set. The log-log plot demonstrates a high degree of linearity, indicating that the model learns properly.\cite{VonLilienfeld2018,Viering2023} The corresponding MAE and Max error values of predictions are shown in SI Fig.~S3. The values decrease by about 35 meV and 160 meV, respectively. The R$^2$ score reaches a maximum of about 1.00 as the training set size increases (SI Fig.~S3b). Notably, with about 1000 data points, the RMSE is already $\sim$35 meV. Further increases in the training set size ($>$1000) result in only marginal improvements in model performance.

To eliminate the possibility of deteriorating model performance due to noise in the data, we further investigate SVR-RBF models trained on separate direct and indirect bandgap datasets via learning curves (SI Figs.~S4a, S4b). As we find that most of the incorrect predictions in the bandgap nature occur in the vicinity of direct-indirect transition (DIT) regions, we additionally remove all data points predicted incorrectly by the classification model and retrain SVR-RBF models with the reduced dataset (SI Fig.~S4c). The figures (SI Fig.~S4) demonstrate that the learning behavior does not change significantly, leading to the conclusion that performance saturation is not due to the severe noise in the data, but rather due to models reaching their maximum capacity to capture the patterns and relationships in the data given.\cite{VonLilienfeld2018,Viering2023} The errors in the bandgap value predictions are spread over the entire composition-strain space. These results indicate that learning both direct and indirect bandgaps simultaneously does not lead to confusion in the SVR-RBF model training.

The prediction accuracy-score of the nature of bandgaps from SVC-RBF models increases by about 12\% as the training set size reaches the largest size (Fig.~\ref{fig:fig1b}). The highest accuracy-score that could be reached is $\sim$0.94 (the last point from Fig.~\ref{fig:fig1b}). The corresponding balanced-accuracy-score values follow the same (SI Fig.~S5). The model performances again saturate at about 1000 data points. These results indicate that the training set is adequate, and around 1000 data points should be sufficient for screening the system being studied.
\begin{table}
  \caption{The out-of-sample bandgap predictions accuracy for the SVM-RBF models from the trial set of the last point from Fig.~\ref{fig:fig1}.}
  \label{table:table1}
    \begin{ruledtabular}
    \begin{tabular}{lll}
    Model & Metric\footnote{Values in brackets state the standard deviation. Error measures (RMSE, MAE, Max error) are given in meV. For MAE and R$^2$, the standard deviations are below 0.5 and 0.005, respectively.} &  \\
    \hline
    \multirow{4}{*}{SVR-RBF} & RMSE & 31 ($\pm$1) \\
    & MAE & 22 ($\pm$0) \\
    & Max error &  155 ($\pm$16)\\ 
    & R$^2$ & 1.00 ($\pm$0.00) \\
    \hline
    \multirow{2}{*}{SVC-RBF} & Accuracy score & 0.94 ($\pm$0.01)\\
    & Balanced accuracy score & 0.94 ($\pm$0.01)\\ 
    \end{tabular}
    \end{ruledtabular}
\end{table}

Moreover, the ML models exhibit excellent efficiency. The process of model training, including hyperparameter optimization, with the largest dataset (3210 training data points) takes only a few minutes on a 6-core CPU.

Table~\ref{table:table1} summarizes the out-of-sample prediction accuracy of the final SVM-RBF models (the last point of Fig.~\ref{fig:fig1}). The RMSE of the bandgap magnitude predictions of the corresponding SVR-RBF models is 31~($\pm$1) meV. For the classification task, the accuracy-score is 0.94~($\pm$0.01). The comparison of all the DFT calculated and the ML-predicted bandgap values and types are presented in SI Figs.~S6 and S7. The final trained models and the list of optimized hyperparameter values of the corresponding models are attached in the SM.

Overall, the errors in bandgap predictions are well within the uncertainty of the most accurate DFT results. We thus use the models from the last point from Fig.~\ref{fig:fig1} to construct the bandgap phase diagram for GaAsPSb over the complete composition range (As, P, Sb = 0--100\% ) and up to $5\%$ strain. 

We emphasize that relying on a simple bowing model for bandgap variation\cite{Vurgaftman2001Band} in such multinary compounds would require an extensive number of bowing parameters for composition and strains. As stated in the introduction, there is a lack of extensive experimental data or bowing parameters available specifically for GaAsPSb. Furthermore, the variation of bandgap values under strain for different III-V compositions often deviates from the simple quadratic dependence expected from a bowing model,\cite{Mondal2022} and they lack information about the bandgap nature. Therefore, the combined accurate DFT-ML hybrid approach developed in the article is necessary for an accurate determination of bandgap values and natures.
\begin{figure}
    \centering
    \includegraphics[height=8.2cm]{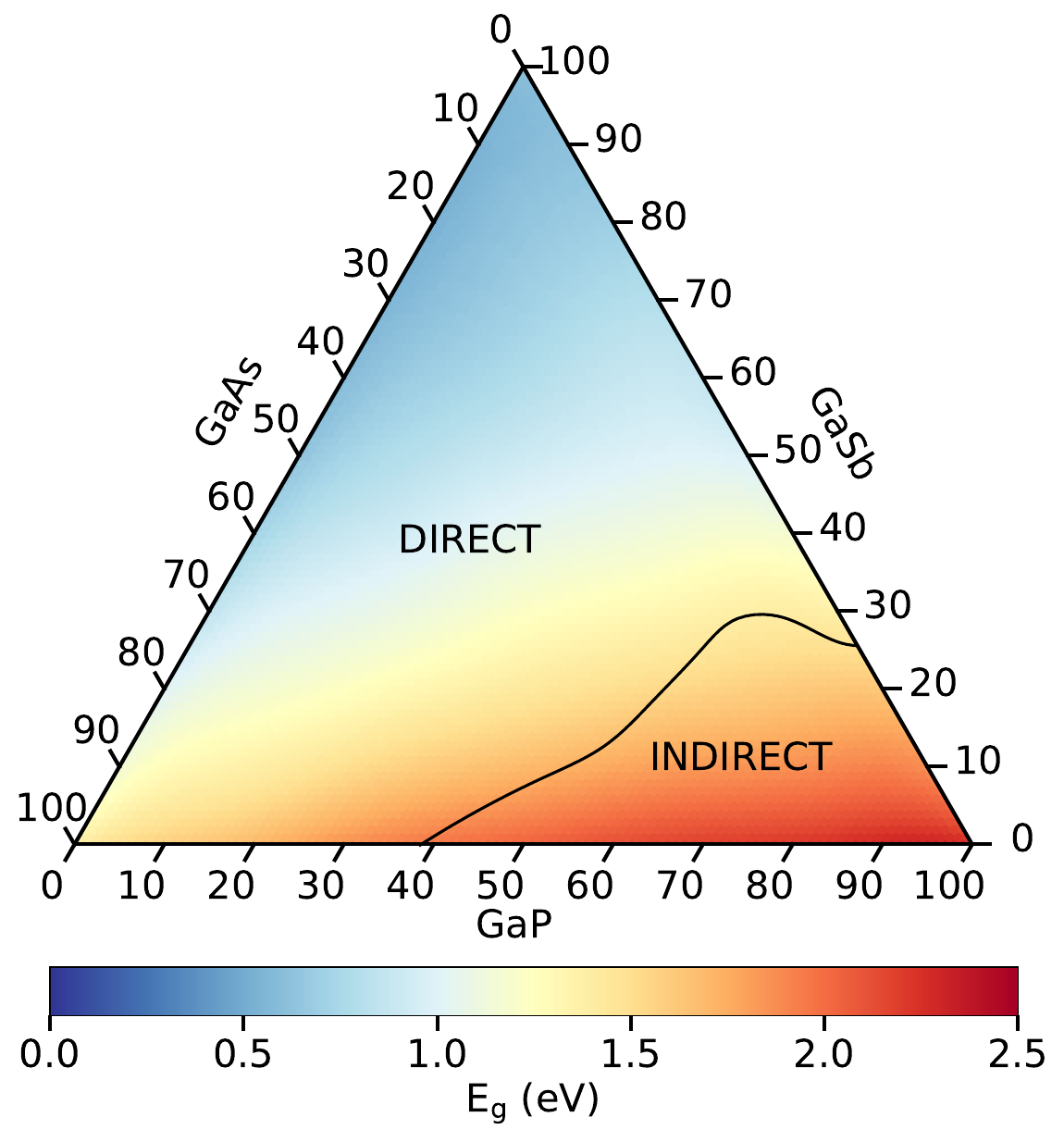}
    \caption{Variation of bandgap values (E$_{\rm g}$) for unstrained GaAsPSb (0.0\% strain). The labels `direct' and `indirect' describe the enclosed regions, with the nature of bandgap being direct and indirect, respectively. The bandgap values are the average values over the 5 model predictions from the trial set of the last point from Fig.~\ref{fig:fig1a}. The nature of the bandgaps are the most frequent outcomes over the 5 predictions (mode value) from the trial set of the last point from Fig.~\ref{fig:fig1b}.}
    \label{fig:fig2}
\end{figure}
\begin{figure}
    \centering
    \includegraphics[height=8.2cm]{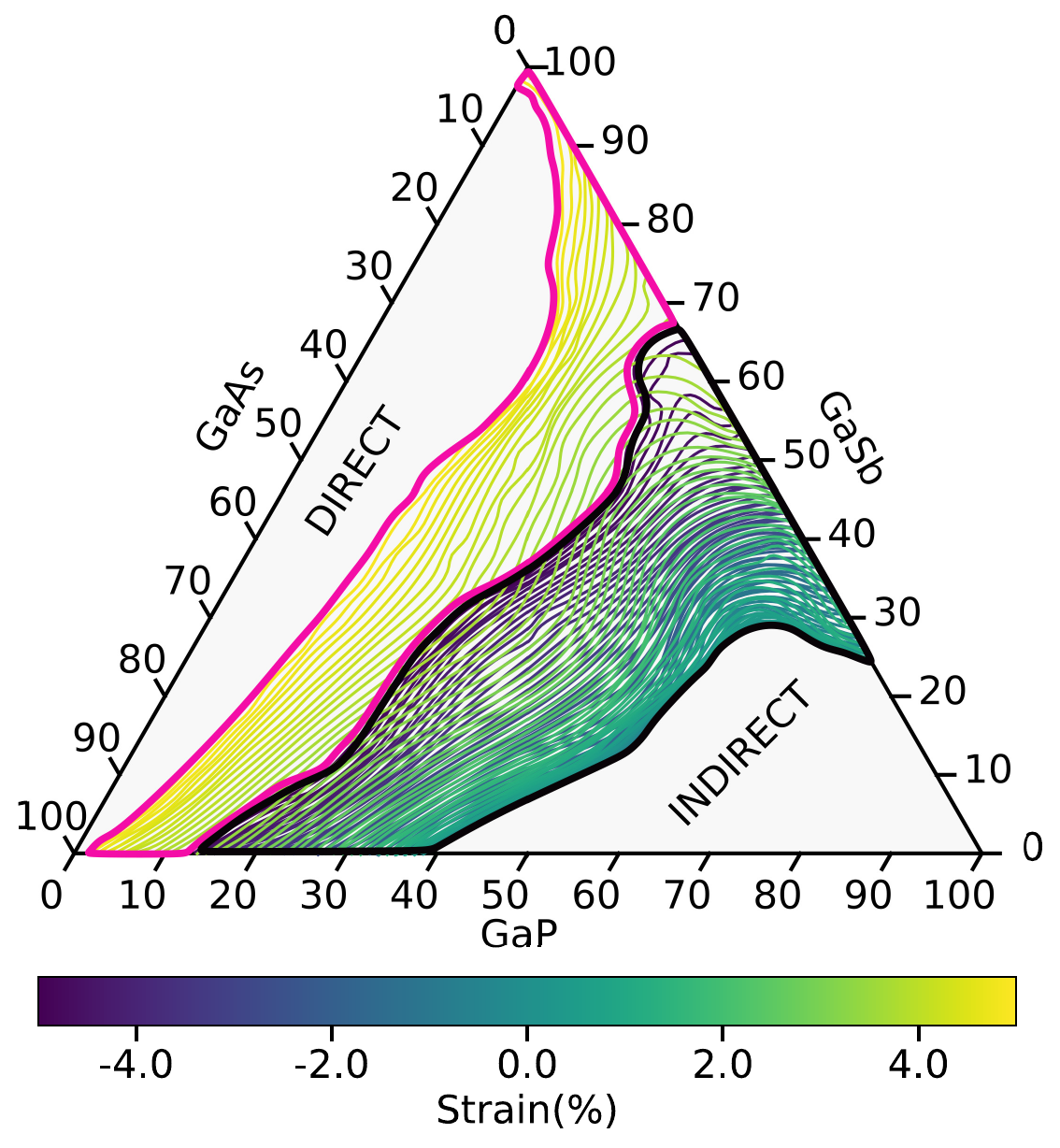}
    \caption{DIT lines for strained GaAsPSb (up to $5\%$ strain). The labels `direct' and `indirect' describe the enclosed regions, with the nature of bandgap being direct and indirect, respectively. The DITs occur under both tensile and compressive strain in the  area enclosed by bold black solid line. In the region enclosed by magenta curve, DITs are possible only under tensile strain. No DIT occurs in the other regions.}
    \label{fig:fig3}
\end{figure} 

\subsection{\label{subsec:gaaspsbbpd}GaAsPSb bandgap phase diagram}
\begin{figure*}
\centering
 \subfloat[]{\label{fig:fig4a}\includegraphics[height=8.2cm]{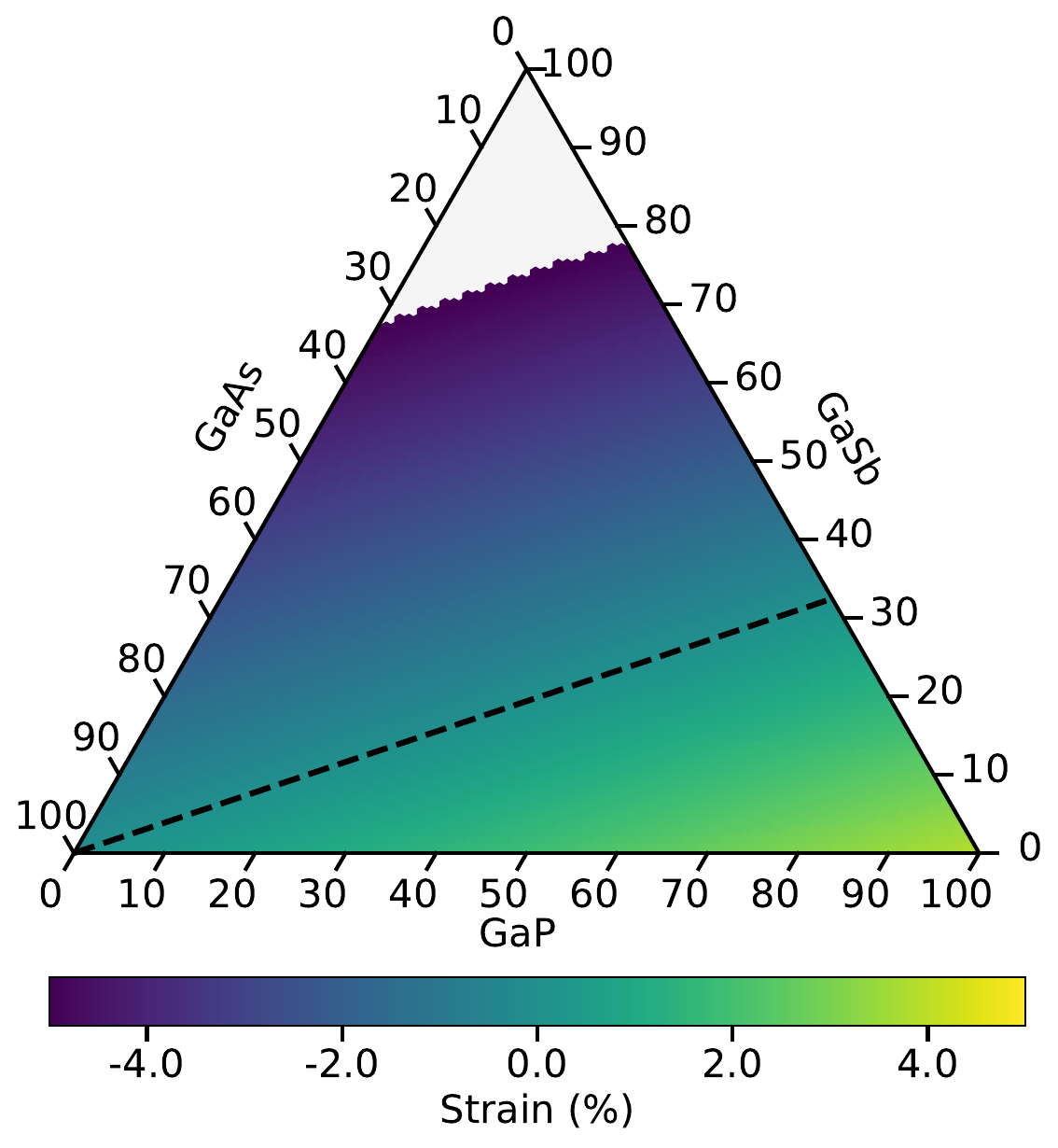}}
 \hspace{1cm}
 \subfloat[]{\label{fig:fig4b}\includegraphics[height=8.2cm]{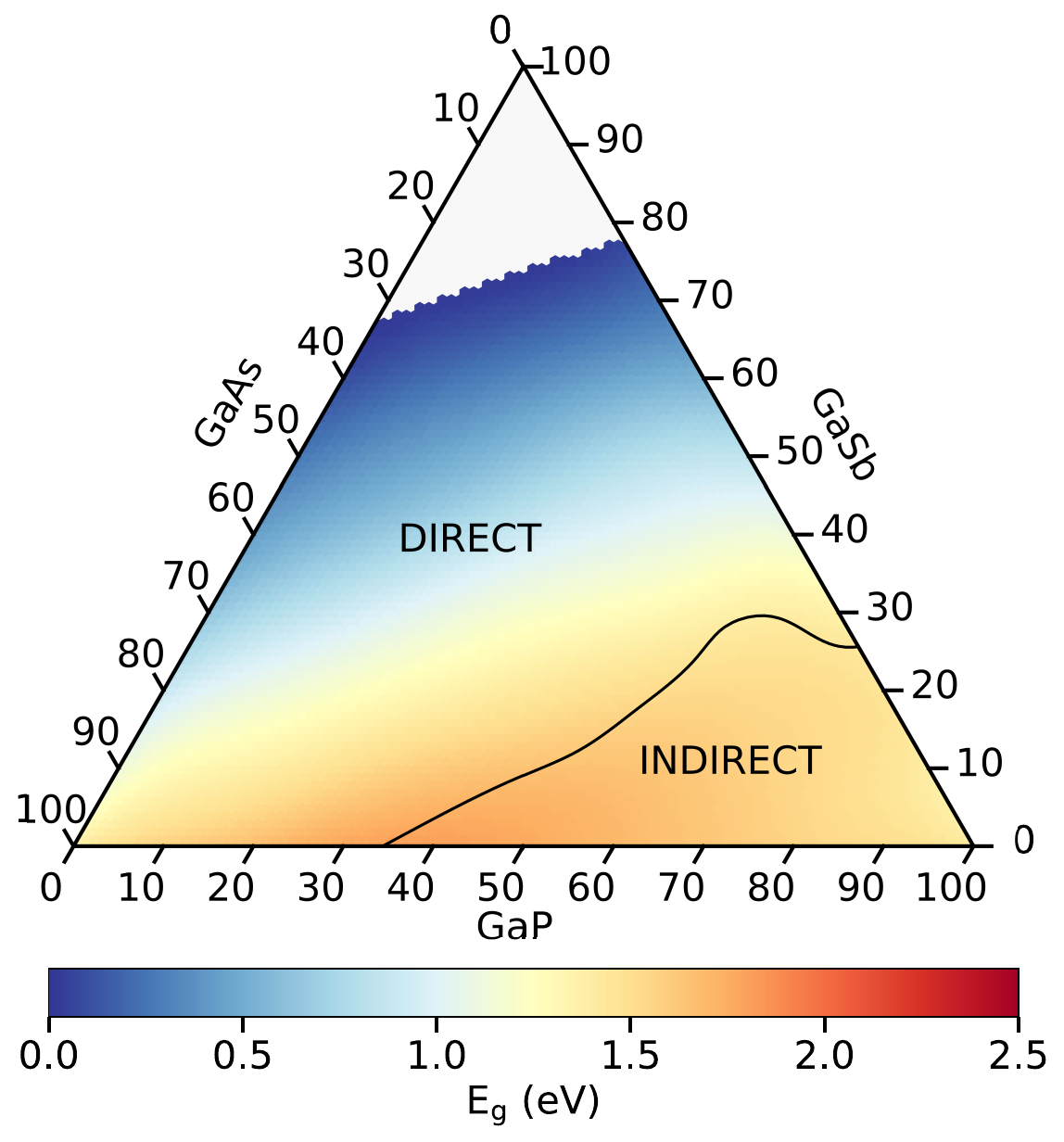}}
   \caption{Effect of GaAs-substrate on GaAsPSb epi-layer under the `theoretical epitaxy' model\cite{Mondal2022,Mondal2023} (up to $5\%$ strain). (a) The calculated biaxial strain values from Eqs.~\ref{eqn:eqn8} and \ref{eqn:eqn9}. The black dotted line indicates the perfect lattice matching (no strain) compositions. (b) The predicted bandgap values in color (E$_\text{g}$). The labels `direct' and `indirect' describe the enclosed regions, with the nature of bandgap being direct and indirect, respectively. The bandgap values are the average values over the 5 model predictions from the trial set of the last point from Fig.~\ref{fig:fig1a}. The nature of the bandgaps are the most frequent outcomes over the 5 predictions (mode value) from the trial set of the last point from Fig.~\ref{fig:fig1b}.}
   \label{fig:fig4}
\end{figure*}
Figure~\ref{fig:fig2} shows the bandgap phase diagram for unstrained GaAsPSb. The bandgap values shown are the average values obtained from the predictions of 5 models. The small standard deviations ($<$35 meV) of the five model predictions (as shown in SI Fig.~S8) indicate that the training set covers the whole phase space sufficiently and that the predictions made across the entire phase space sampled are accurate. In addition, the use of multiple models for prediction increases the robustness of the predictions of regression models by a factor of the square root of the number of models,~\cite{Gastegger2017} making predictions in general more precise and less noisy. For the nature of bandgaps, we use the most frequent outcomes over the 5 predictions (mode value). The labels `direct' and `indirect' in the figure indicate the area where the nature of bandgaps is direct and indirect, respectively. These areas are separated by the DIT line (black curve). We smoothen the calculated DIT points with B-spline function~\cite{Dierckx1982} (for details, see SI Sec.~S7). The complete movie of the corresponding diagrams for other strain values is available in the SM attachment. Figure~\ref{fig:fig2} shows that the largest bandgaps, which are indirect in nature, are found for high \%P. The bandgaps become smaller as more \%As and \%Sb are added. The bandgap values also decrease with increase in tensile as well as compressive  strain (SI Fig.~S10). 

Further, we group the composition space DIT curves for all the strain values investigated, as shown in Fig.~\ref{fig:fig3}. We find that within the strain regime investigated here, in the area indicated by the magenta box in Fig.~\ref{fig:fig3}, the DITs are only possible under tensile strain. In the area indicated by the bold black box, the DITs can be achieved using both tensile and compressive strain. No DITs are found in other regions.

\subsection{\label{subsec:subeffectongaaspsbbandgap}Modeling GaAsPSb epitaxy}
One of the most common approaches to experimentally realize quaternary III-V semiconductors is epitaxial growth. As pointed out in Ref.~\citenum{Mondal2023}, biaxial strain can be used to model epitaxial growth (`theoretical epitaxy'). We thus also investigate the effect of different substrates. We first calculate the equilibrium lattice parameters of the unstrained quaternary compositions using Vegard's law,\cite{Vegard1921,Ashcroft1991} Eq.~\ref{eqn:eqn8}. 
\begin{align}
     a_{_{\text{GaAs}_x\text{P}_{y}\text{Sb}_{100-x-y}}} = \,&\frac{1}{100}[x\times a_{_{\text{GaAs}}} +y\times a_{_{\text{GaP}}} \nonumber\\
     &+ (100-x-y)\times a_{_{\text{GaSb}}}] \label{eqn:eqn8}
\end{align}
with $a_{_{\text{GaP}}} = 5.475\,\text{\AA}, a_{_{\text{GaAs}}} = 5.689\,\text{\AA}, a_{_{\text{GaSb}}} = 6.134\,\text{\AA}$.\cite{Mondal2022}\\ \\
The biaxial strain values are then computed using Eq.~\ref{eqn:eqn9}.
\begin{equation}
    \text{Strain (\%)} = \frac{a_{_{\text{substrate}}} - a_{_{\text{GaAsPSb}}}}{a_{_{\text{GaAsPSb}}}}\times 100 \label{eqn:eqn9}
\end{equation}
Figure~\ref{fig:fig4a} shows the strain map when the respective GaAsPSb compositions are to be grown on GaAs substrate.

We then predict the bandgaps using the SVM-RBF models described earlier (see Fig.~\ref{fig:fig4b}). In the region with high Sb content, we find large compressive strain. The bandgap values in those regions are very small. On the contrary, in the high \%P region, we find large tensile strain. The bandgaps there are indirect in nature and thus are not appropriate for optoelectronic applications. The perfect lattice matching condition can be found in between, as shown by the dotted line in Fig.~\ref{fig:fig4a}.

From the above discussions, it becomes clear that one can optimize the best material combinations for achieving specific optical applications using these diagrams. Similar diagrams for other substrates (GaP, GaSb, InP, and Si) can be found in SI Fig.~S11. We also provide interactive figures of the above GaAsPSb bandgap phase diagrams in the SM attachment.

Further, for the largest training dataset (the last point of Fig.~\ref{fig:fig1}) we predict the bandgap nature using SVC-RBF model with the optimized hyperparameters from the SVR-RBF models. We obtain an accuracy score of 0.94~($\pm$0.01) in the out-of-sample performance. This is comparable to the performance of the hyperparameter-optimized SVC-RBF model (Table~\ref{table:table1}). Similarly, we employ the SVR-RBF model with optimized hyperparameters from the SVC-RBF models to predict the bandgap magnitudes. For all the 5 SVR-RBF models here, we set the $\varepsilon$ value to 0.02, which was the average value of previously hyperparameter-optimized SVR-RBF models. This results in an out-of-sample RMSE of 32~($\pm$1) meV for the predictions, again demonstrating accuracy similar to that of the optimized SVR-RBF models (Table~\ref{table:table1}). This suggests that separate hyperparameter optimizations are not required for SVR-RBF and SVC-RBF models independently. The optimized hyperparameters from the SVR-RBF model can be used for the prediction of SVC-RBF and vice-versa.

\section{\label{sec:summary}Summary}
In this work, using machine learning models trained on first-principles calculations, we developed an efficient DFT-ML hybrid computational approach for mapping the bandgap magnitude and type (direct or indirect) over a wide range of composition and strain values in quaternary III-V semiconductors. The devised SVR-RBF and SVC-RBF-based ML models showed remarkable accuracy in predicting the bandgap properties in these materials despite using a very simple descriptor, i.e., composition and strain values. This strategy significantly accelerates the sampling efficiency over the large strain-composition space in multinary systems, which otherwise would be impossible to cover with first-principle calculations only. Using the protocol, we constructed the composition-strain-bandgap mapping, the bandgap phase diagram, for GaAsPSb. This diagram can be used as a valuable tool for making judicial choices of best materials to target applications. The study revealed that within 5\% strain, GaAsPSb compositions with high P concentration are characterized by indirect bandgaps and are thus not suitable for optoelectronic devices. Notably, our computational protocol can be generalized to other multinary III-V semiconductors. Thus, the rapid estimation of bandgaps for a large number of composition and strain values using this approach will be extremely useful for screening multinary III-V materials. This provides a powerful approach for future materials design, facilitating the development of novel strain-engineered semiconductor materials with tailored bandgap properties.

\section*{Supplementary material}
Supplementary Materials include:
\begin{itemize}
    \item Supporting\_Information.pdf: collection of supplementary supporting figures and tables.
    \item GaAsPSb\_ML\_database.zip: complete dataset for ML training and testing, contains \\``GaAsPSb\_ML\_database.xlsx''.
    \item GaAsPSb\_bandgap\_phase\_diagram.mp4: bandgap phase diagram movie for GaAsPSb up to $5\%$ tensile and compressive strain.
    \item Interactive\_bandgap\_phase\_diagram.zip: interactive diagrams in HTML format. To view the diagrams, open the HTML files in a web browser.
    \begin{itemize}
        \item Substrate\_effect\_strain\_map.html: biaxial strain map of GaAsPSb when grown on different substrates (calculated using Eqs.~\ref{eqn:eqn8} and \ref{eqn:eqn9}) .
        \item Substrate\_effect\_bandgap\_phase\_diagram.html: ML predicted bandgap map of GaAsPSb when grown on different substrates.
        \item Direct\_indirect\_transition\_lines.html: ML predicted direct-indirect transition lines (without B-spline fitting) for GaAsPSb up to $5\%$ tensile and compressive strain. 
        \item Bandgap\_phase\_diagram\_GaAsPSb.html: bandgap phase diagram for GaAsPSb (within $5\%$ tensile and compressive strain). The direct-indirect transition lines are without B-spline fitting.
    \end{itemize}
    \item GaAsPSb\_MachineLearning\_scripts.zip: Python scripts for the ML codes used to generate the data in this manuscript. Additionally, a user guide titled ``Helpers.txt'' is included within the zip file to assist with executing the ML scripts.
    \item FinalBest5TrainedModels.zip: final best five ML trained models used to create the bandgap phase diagram in this manuscript. The zip file includes a Python script that can be used to predict the bandgaps of GaAsPSb system and a user guide titled ``Helpers.txt'' to assist with executing the script.
    \item FinalBest5TrainedModels\_hyperparameters.zip: list of optimized hyperparameter values of the final best five SVR-RBF and SVC-RBF models, contains \\``FinalBest5TrainedModels\_hyperparameters.xlsx''.
\end{itemize}

\begin{acknowledgements}
We acknowledge German Research Foundation (DFG) in the framework of the Research Training Group ``Functionalization of Semiconductors'' (GRK 1782) for funding this project and to HRZ Marburg, GOETHE-CSC Frankfurt, ZIH TU Dresden, and HLR Stuttgart for providing the computational resources. We thank Dr. Marcel Kr{\"o}ner and Prof. Dr. Kerstin Volz from Philipps-Universit{\"a}t Marburg for discussions.
\end{acknowledgements}
\section*{Author declarations}
\subsection*{Conflict of interest}
The authors have no conflicts to disclose.
\subsection*{Author contributions}
BM and RTZ contributed to the conception and design of the study, while BM was responsible for performing the calculations and analysis. JW and RTZ provided critical feedback throughout the project. BM drafted the manuscript, which was then reviewed and revised by JW and RTZ. All authors reviewed and approved the manuscript prior to submission.
\section*{Data availability}
The density functional theory calculation data that are used in this manuscripts (see Sec.~\ref{sec:computationaldetails}) are available in the following NOMAD repositories:
\begin{itemize}
    \item B. Mondal and R. Tonner-Zech (2022). \\``III-V binary semiconductors strain study,'' NOMAD. \\ {\small \url{https://doi.org/10.17172/NOMAD/2022.08.20-2}}
    \item B. Mondal and R. Tonner-Zech (2023). \\``III-V ternary semiconductors strain study,'' NOMAD. \\ {\small\url{https://doi.org/10.17172/NOMAD/2023.02.27-1}}
    \item B. Mondal, J. Westermayr, and R. Tonner-Zech (2023). ``GaAsPSb bandgap phase diagram,'' NOMAD. \\ {\small\url{https://doi.org/10.17172/NOMAD/2023.05.03-1}}.
\end{itemize}

The python scripts implementing the ML models used in this manuscript can be found in the SM attachment (GaAsPSb\_MachineLearning\_scripts.zip file). The final ML-trained models used to construct the bandgap phase diagrams are also attached to the SM (FinalBest5TrainedModels.zip file).   

In addition to the SM attachments, the interactive bandgap phase diagrams can also directly be viewed at \url{https://bmondal94.github.io/Bandgap-Phase-Diagram/}.\cite{bandgap_phase_diagram_github}

\section*{REFERENCES}
\bibliographystyle{aipnum4-1}
\providecommand{\noopsort}[1]{}\providecommand{\singleletter}[1]{#1}

\end{document}


\title{Supporting Information \\ \vspace{1cm} Machine learning for accelerated bandgap prediction \\in strain-engineered quaternary III-V semiconductors}
\keywords{}
\author{Badal Mondal}
\affiliation{Wilhelm-Ostwald-Institut f\"ur Physikalische und Theoretische Chemie, Universit\"at Leipzig, 04103 Leipzig, Germany}%
\affiliation{Fachbereich Physik, Philipps-Universit\"at Marburg, 35032 Marburg, Germany}

\author{Julia Westermayr}%
\affiliation{Wilhelm-Ostwald-Institut f\"ur Physikalische und Theoretische Chemie, Universit\"at Leipzig, 04103 Leipzig, Germany}%
\affiliation{Center for Scalable Data Analytics and Artificial Intelligence, Dresden/Leipzig, Germany}%

\author{Ralf Tonner-Zech}
\email{ralf.tonner@uni-leipzig.de}
\affiliation{Wilhelm-Ostwald-Institut f\"ur Physikalische und Theoretische Chemie, Universit\"at Leipzig, 04103 Leipzig, Germany}%

\date{\today}

\preprint{APS/123-QED}
\maketitle

\onecolumngrid

\begin{table}[!htbp]
    \centering
    \begin{tabular}{lcl}
    ML & &: Machine learning \\
    SVM-RBF model & &: Support Vector Machine with Radial Basis Function kernel machine learning model  \\
    SVR-RBF model & &: Support Vector Regression with Radial Basis Function kernel machine learning model \\
    SVC-RBF model & &: Support Vector Classification with Radial Basis Function kernel machine learning model \\
    RMSE & &: Root Mean Squared Error \\
    MAE & &: Mean Absolute Error \\
    Max error & &: Maximum error \\
    R$^2$ & &: Coefficient of determination
    \end{tabular}
\end{table}
\pagebreak
\section{\label{sec:secS1}Hyperparameter optimization}
\begin{figure}[h!]
    \centering
    \subfloat[$\varepsilon=10^{-3}-10^{2}$]{\label{fig:figS1a}\includegraphics[width=3.4in]{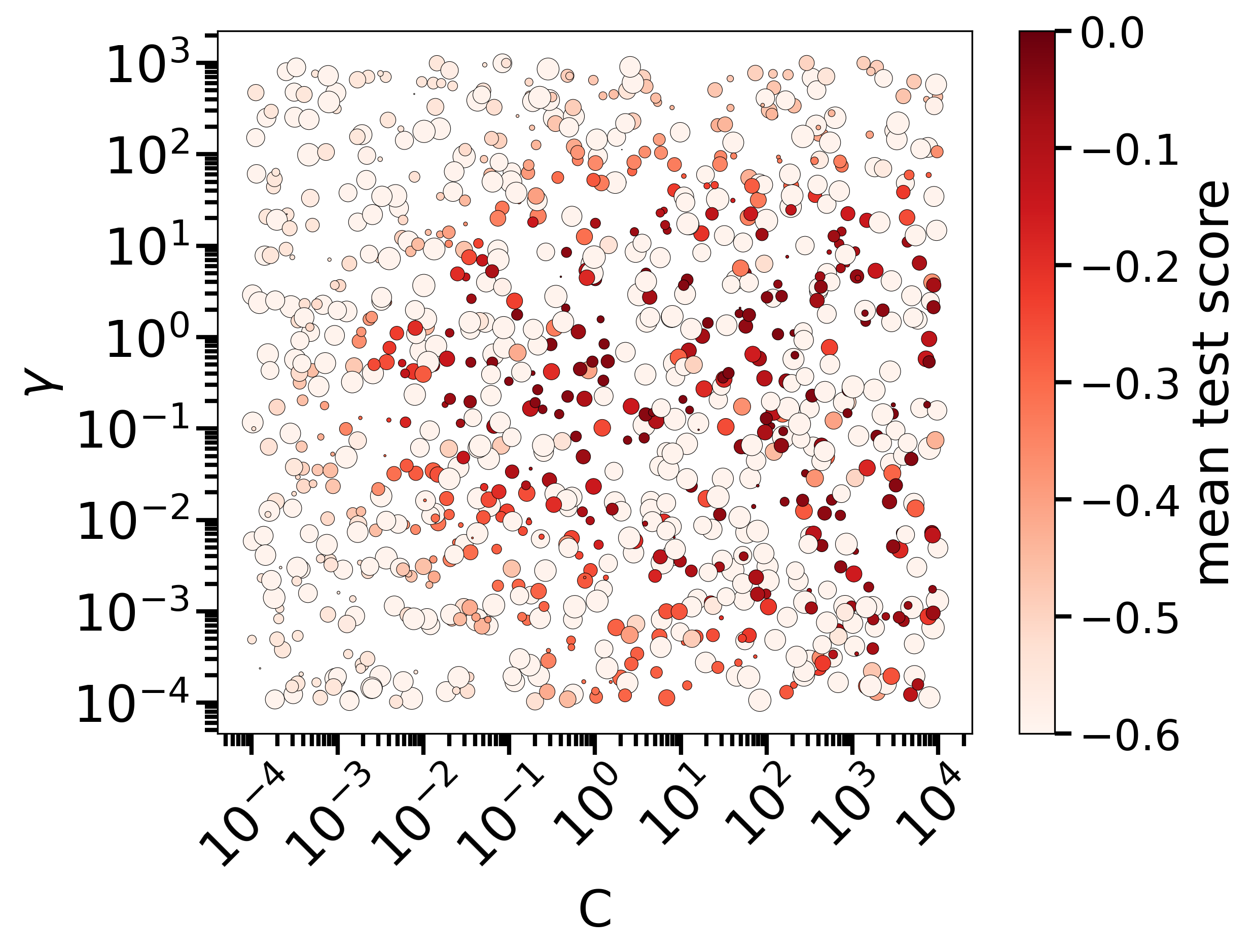}}
    \hspace{.2in}
    \subfloat[$\varepsilon=10^{-2}-10^{2}$]{\label{fig:figS1b}\includegraphics[width=3.4in]{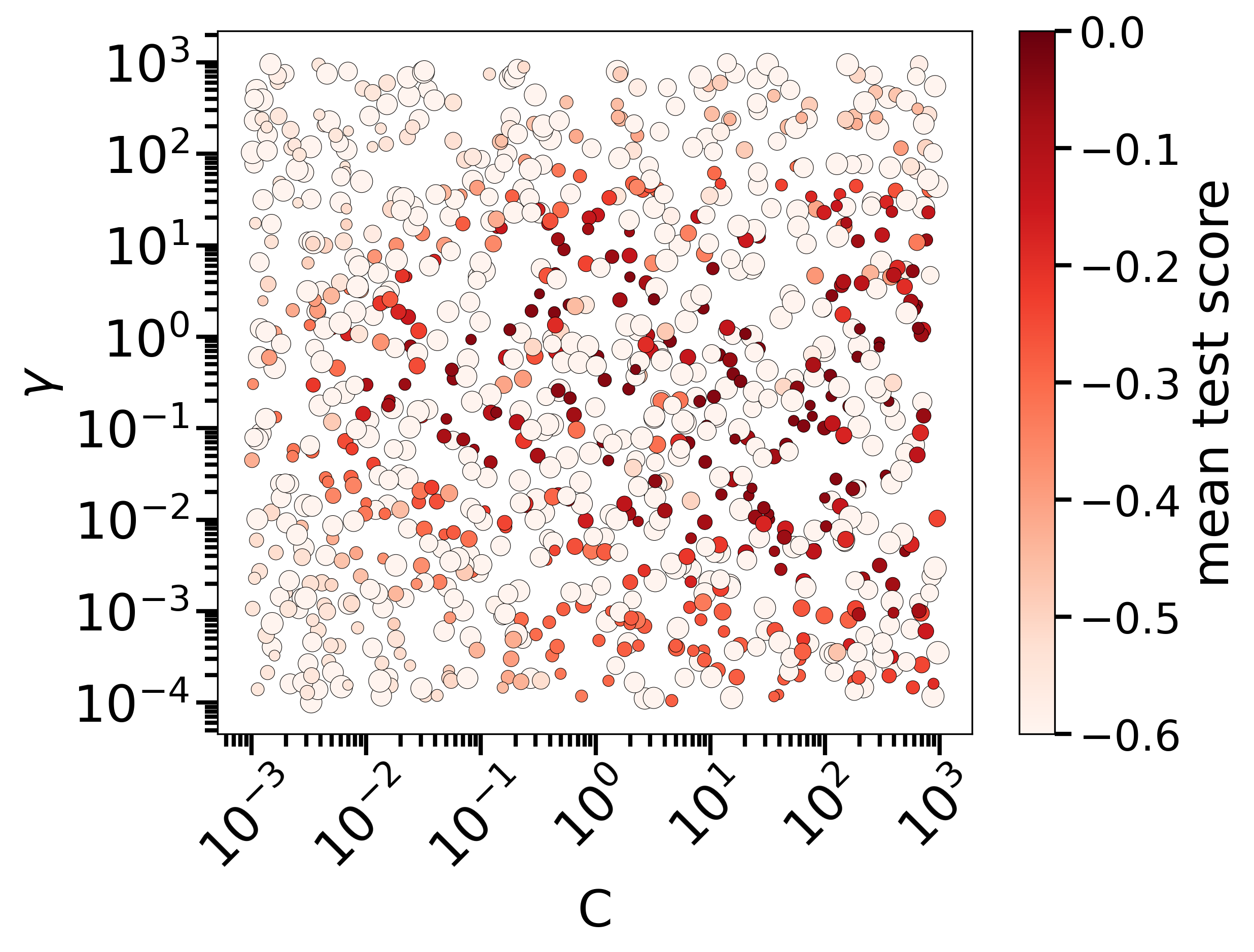}}\\
    \subfloat[$\varepsilon=10^{-2}-10$]{\label{fig:figS1c}\includegraphics[width=3.4in]{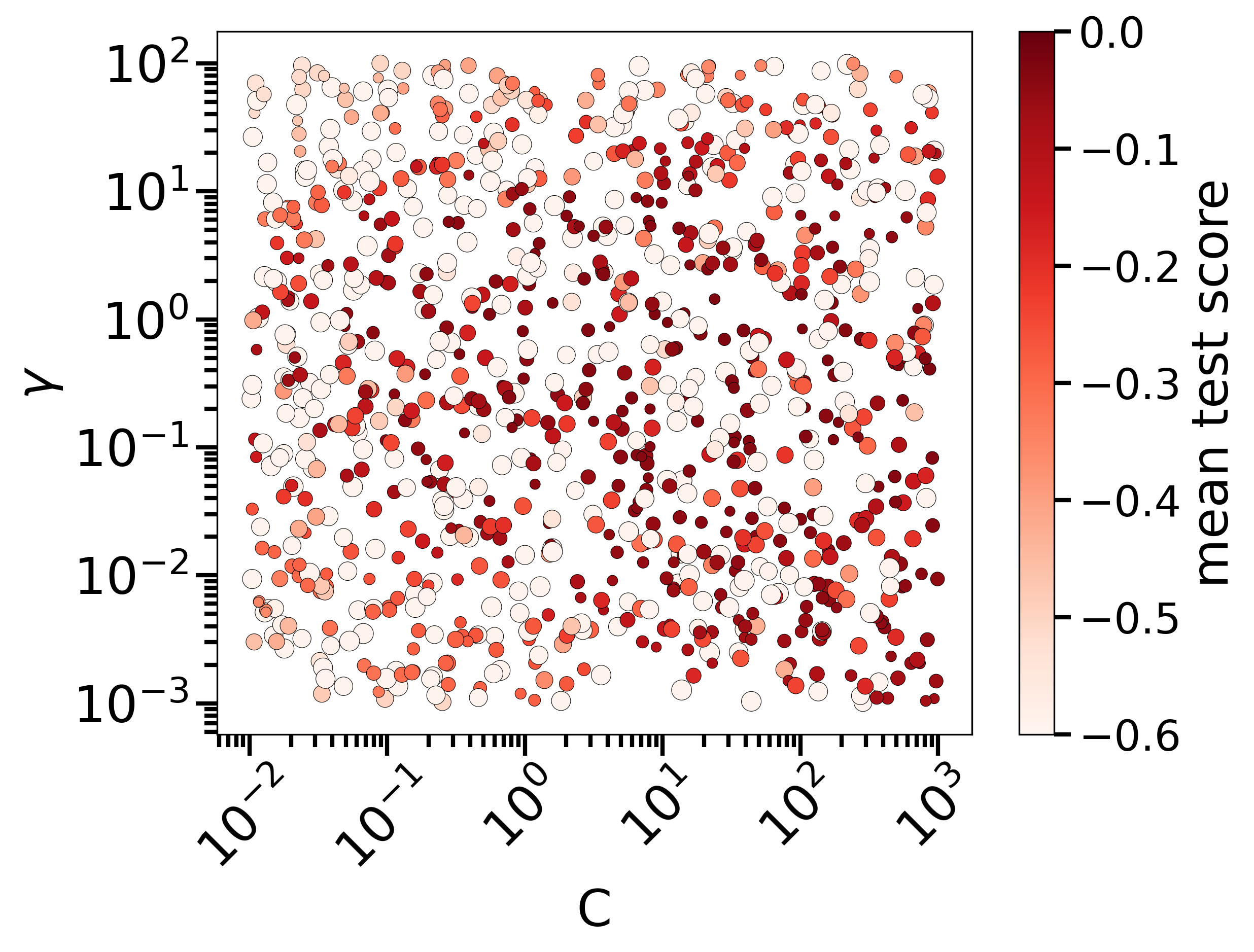}}
    \hspace{.2in}
    \subfloat[$\varepsilon=10^{-2}-1$]{\label{fig:figS1d}\includegraphics[width=3.4in]{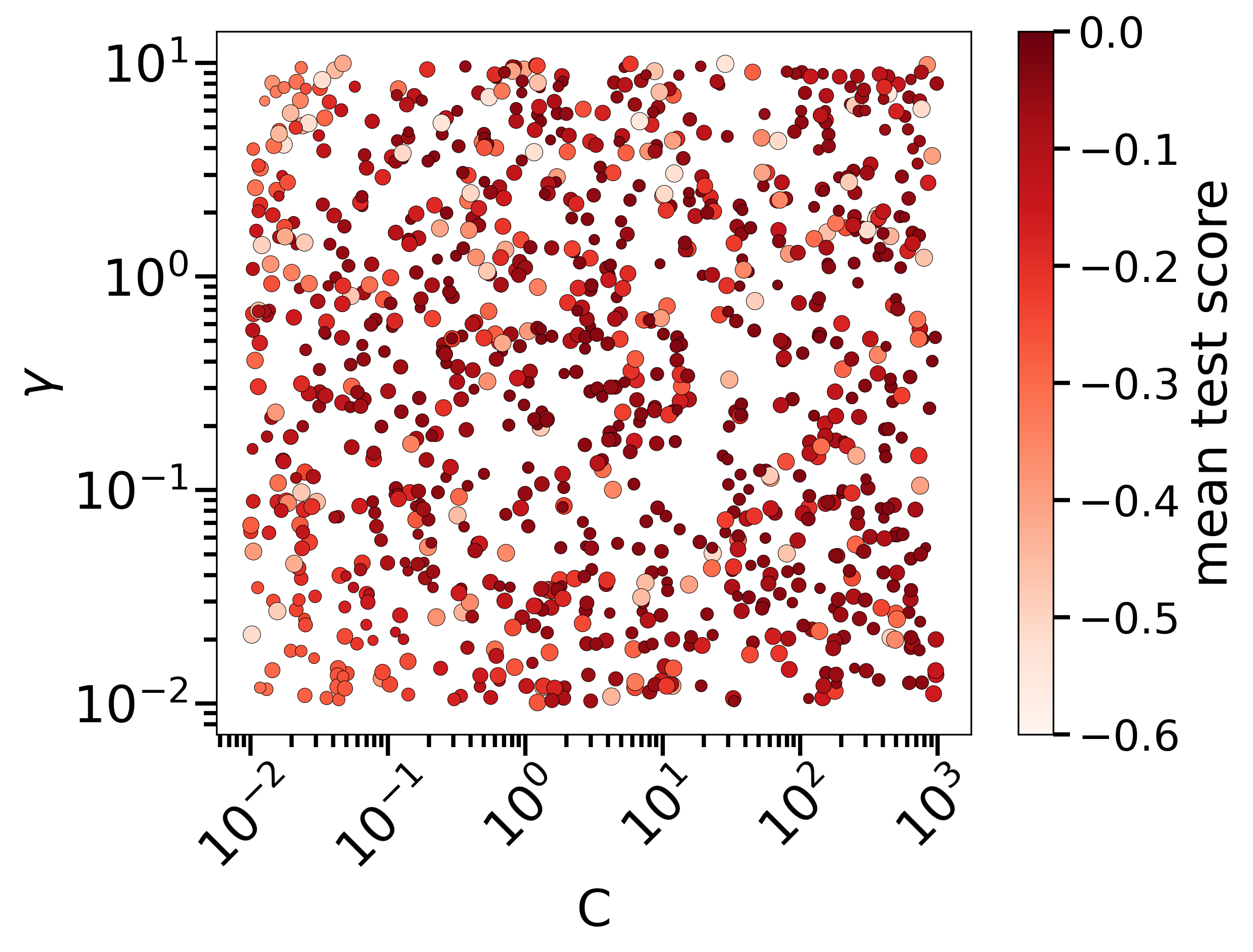}}
    \caption{Hyperparameter optimization with the largest dataset (3210 training data points) using random search cross-validation of the SVR-RBF model. The hyperparameter optimization scoring function is the negative RMSE, meaning that a small error or a high test-score (red) is beneficial. The random search is performed for each subfigure over 1000 random $C$, $\gamma$, and $\varepsilon$ combinations from a logarithmic uniform distribution space. The sizes of the circles, depicted in $\log$-scale, represent the
    $\varepsilon$ values. Specifically, we plot $\log(10^3\varepsilon)$ values to ensure non-negative marker size. The ranges for $C$, $\gamma$, and $\varepsilon$ are made tighter from (a) to (d).}
    \label{fig:figS1}
\end{figure}
In Fig.~\ref{fig:figS1}, we show the hyperparameter optimization with random search cross-validation of the SVR-RBF model. We tune the $C$, $\gamma$, and $\varepsilon$ ranges and find Fig.~\ref{fig:figS1d} as the optimal choice, covering the regimes with the best hyperparameters set. The larger $C$ value ($>10^3$) would result in a hard margin in the SVR model and, thus, poor generalization on unknown data. Therefore, for $C$, no larger values are sampled here. 

Note the mean test score values in Fig.~\ref{fig:figS1} are displayed on a negative scale. As mentioned in the manuscript, we used the package \verb|sklearn| for implementing the ML models in this article. The hyperparameter optimizations in this package are performed by evaluating the scoring function values over the cross-validation set. In \verb|sklearn|, by convention, the convention is to maximize all scorer objects during optimization. However, in ML problems, the objective is to minimize the error for error functions. Thus metrics such as `mean squared error' are available as `negative mean squared error' in hyperparameter optimization scoring functions, which return the negated value of the metric.

\clearpage
\section{\label{sec:secS2}Negative direct bandgap}
\begin{figure}[h!]
    \centering
    \includegraphics[width=3.in]{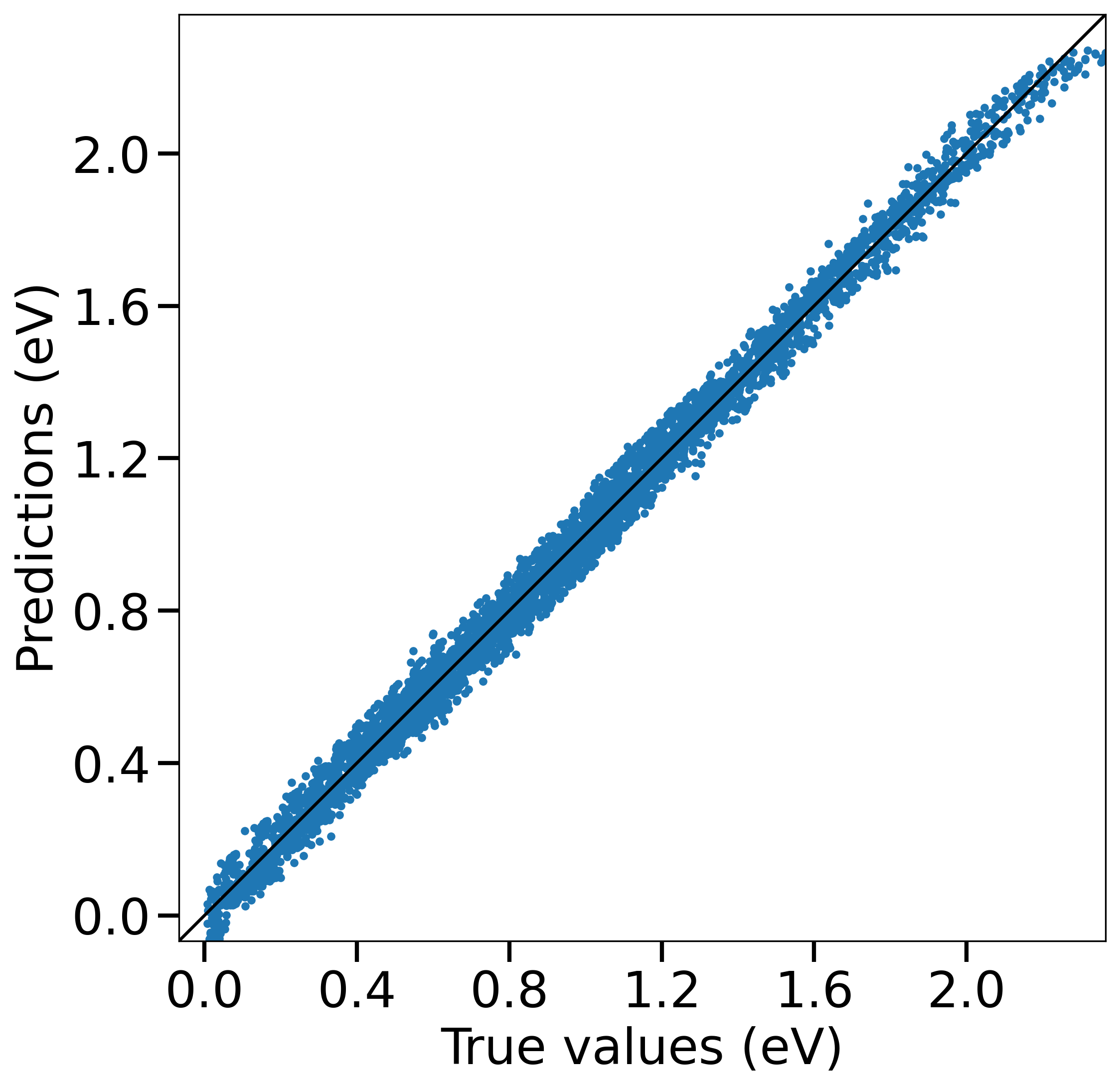}
    \caption{The SVR-RBF model bandgap value predictions over all the samples. The true values are the DFT calculated bandgaps. The hyperparameters are optimized with RMSE metric.}
    \label{fig:figS2}
\end{figure}
In Fig.~\ref{fig:figS2}, the bandgap values were predicted using SVR-RBF model. The hyperparameters of the model were optimized using RMSE metric, Eq.~\ref{eqn:eqnS1}. However, the corresponding SVR-RBF model predicted a few small negative direct bandgap values of up to --5 meV (see left-bottom corner in Fig.~\ref{fig:figS2}). In the manuscript, we thus converted all the predicted negative bandgap values ($\hat{y}_i$) to 0 ($\hat{y}_i$=0). The RMSE of the model predictions was calculated after the conversion. Accordingly, all the performance metrics evaluated did include this conversion.

\begin{eqnarray}
\text{RMSE} = \sqrt{\frac{1}{n_{_{\text{samples}}}}\sum_{i=1}^{n_{_{\text{samples}}}} \left(y_i-\hat{y}_i\right)^2} \label{eqn:eqnS1}
\end{eqnarray}
Where $\hat{y}_i$ is the ML model predicted value of $i$-th sample and $y_i$ is the corresponding true value. $n_{_{\text{samples}}}$ is the total number of samples. 

\section{\label{sec:secS3}Machine learning dataset features and labels for G\lowercase{a}A\lowercase{s}PS\lowercase{b}}
\noindent The full data dataset can be found in the Supplementary Information attachment. In the table below, the ML features for 3 example data are given.\\
Sample 1: Ga$_{100}$P$_0$As$_{100}$Sb$_0$ ($\equiv$ GaAs), unstrained\\
Sample 2: Ga$_{100}$P$_{25}$As$_{25}$Sb$_{50}$, 3.0\% biaxial tensile strained \\
Sample 3: Ga$_{100}$P$_{50}$As$_{50}$Sb$_{0}$, 5.0\% biaxial compressively strained
\begin{table}[h]
\begin{ruledtabular}
\begin{tabular}{c|cccc|cc}
\multirow{2}{*}{Sample index} & \multicolumn{4}{c|}{Features} &  \multicolumn{2}{c}{Labels}\\ \nmidrule{2-7}
 & Phosphorus (\%) & Arsenic (\%) & Antimony (\%) & Strain (\%) & Bandgap value (eV) & Bandgap nature\footnote{The direct and indirect bandgap natures are feature transformed to 1 and 0s before ML training.} \\ \nhrule
 1 & 0 & 100 & 0 & 0.0 & 1.466 & direct \\
 2 & 25 & 25 & 50 & 3.0 & 0.629 & direct \\
 3 & 50 & 50 & 0 & -5.0 & 1.243 & indirect 
\end{tabular}
\end{ruledtabular}
\end{table}

\clearpage
\section{\label{sec:secS4}Dataset convergence}
\begin{figure}[!h]
\centering
  \subfloat[]{\label{fig:figS3a}\includegraphics[width=2.8in]{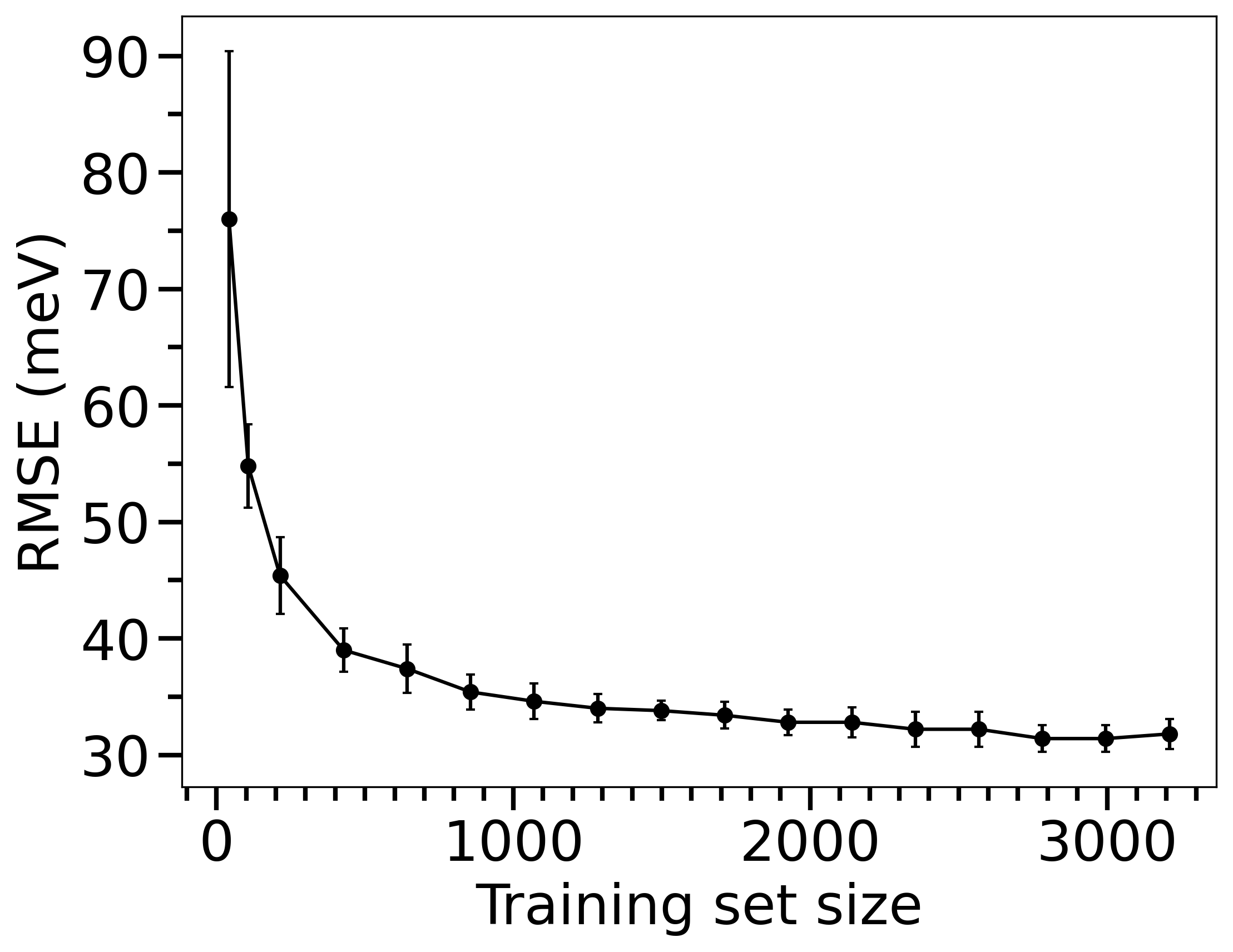}}
  \hspace{.3in}
  \subfloat[]{\label{fig:figS3b}\includegraphics[width=3.in]{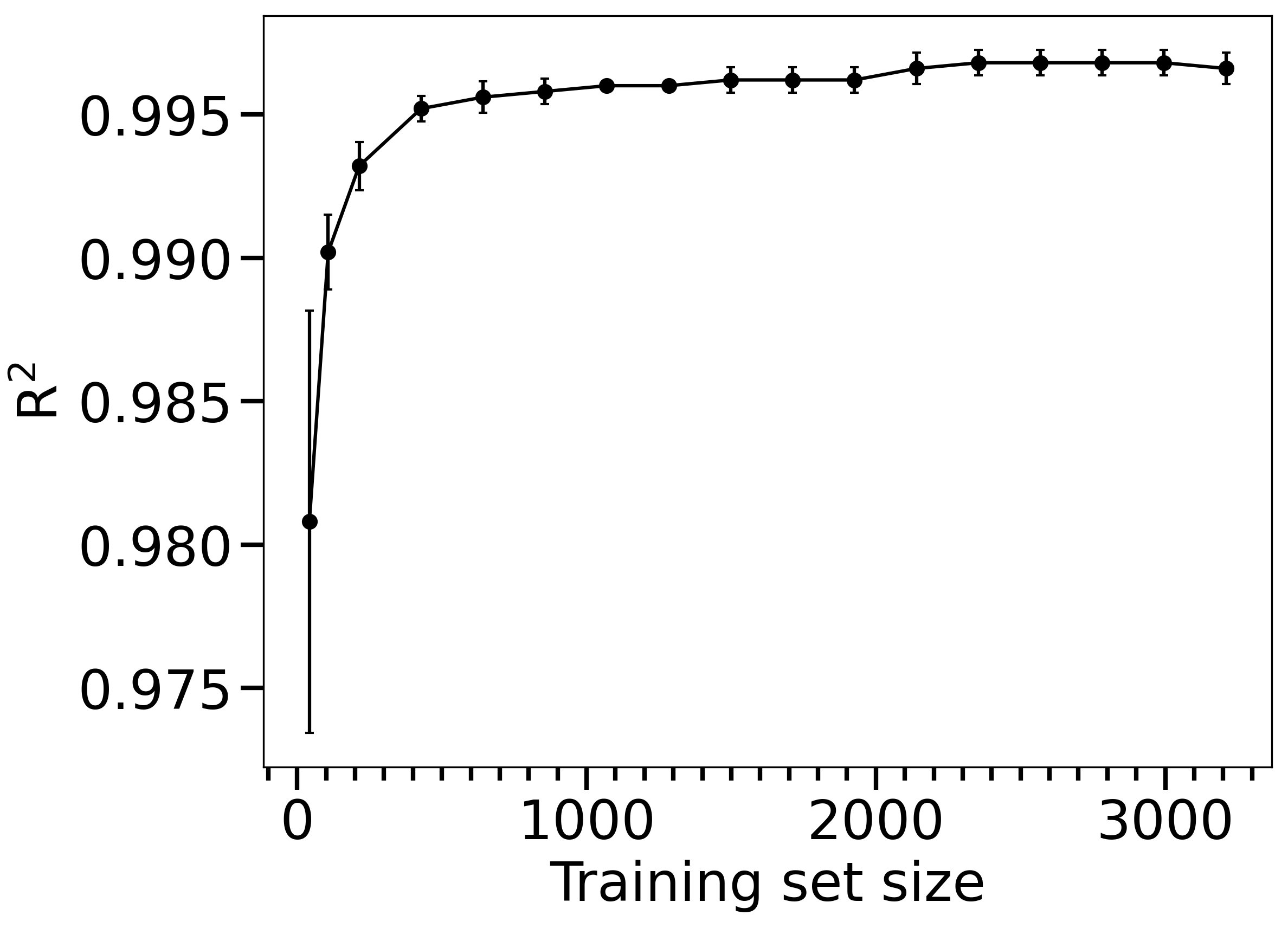}}\\
  \subfloat[]{\label{fig:figS3c}\includegraphics[width=3.in]{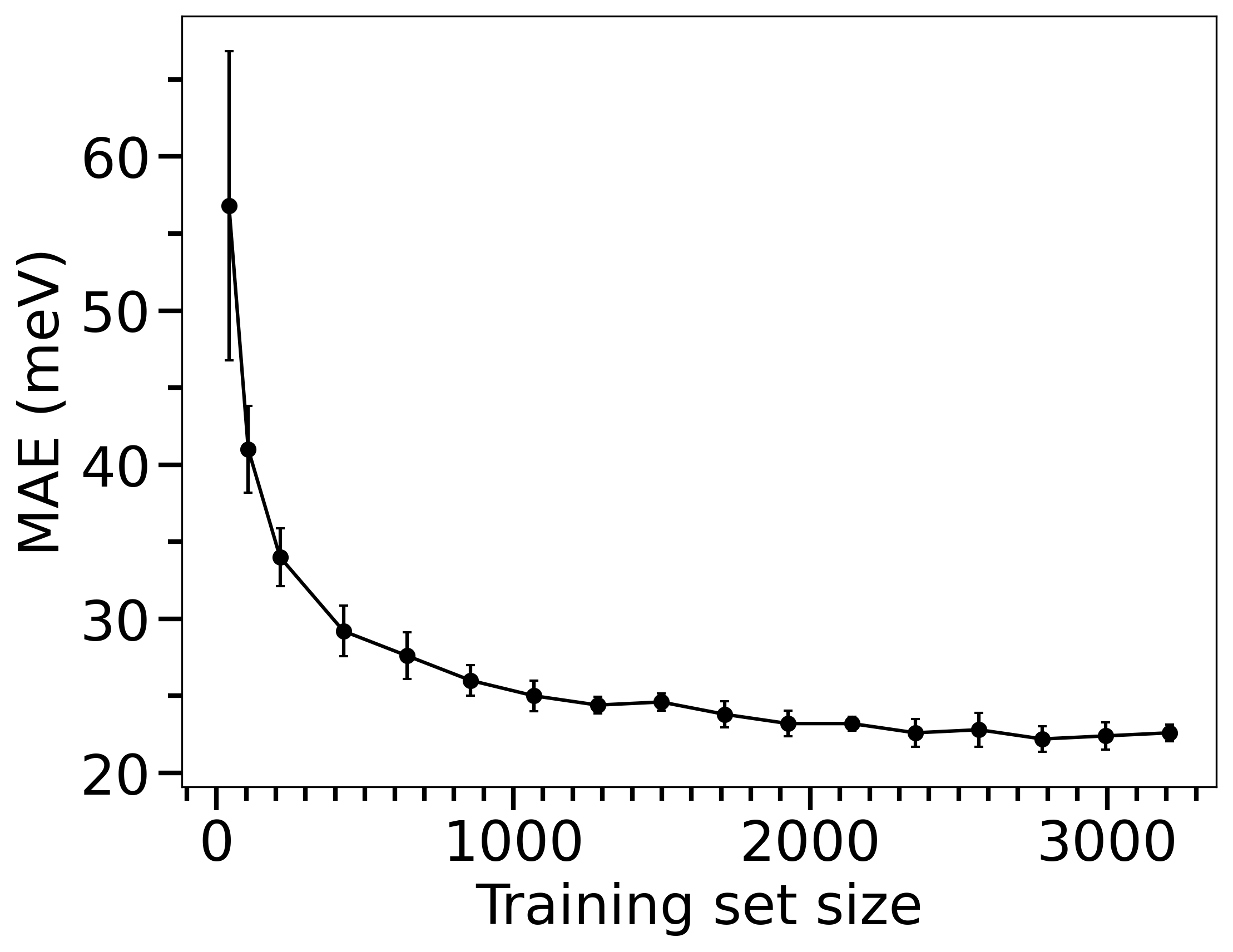}}
  \hspace{.3in}
  \subfloat[]{\label{fig:figS3d}\includegraphics[width=3.in]{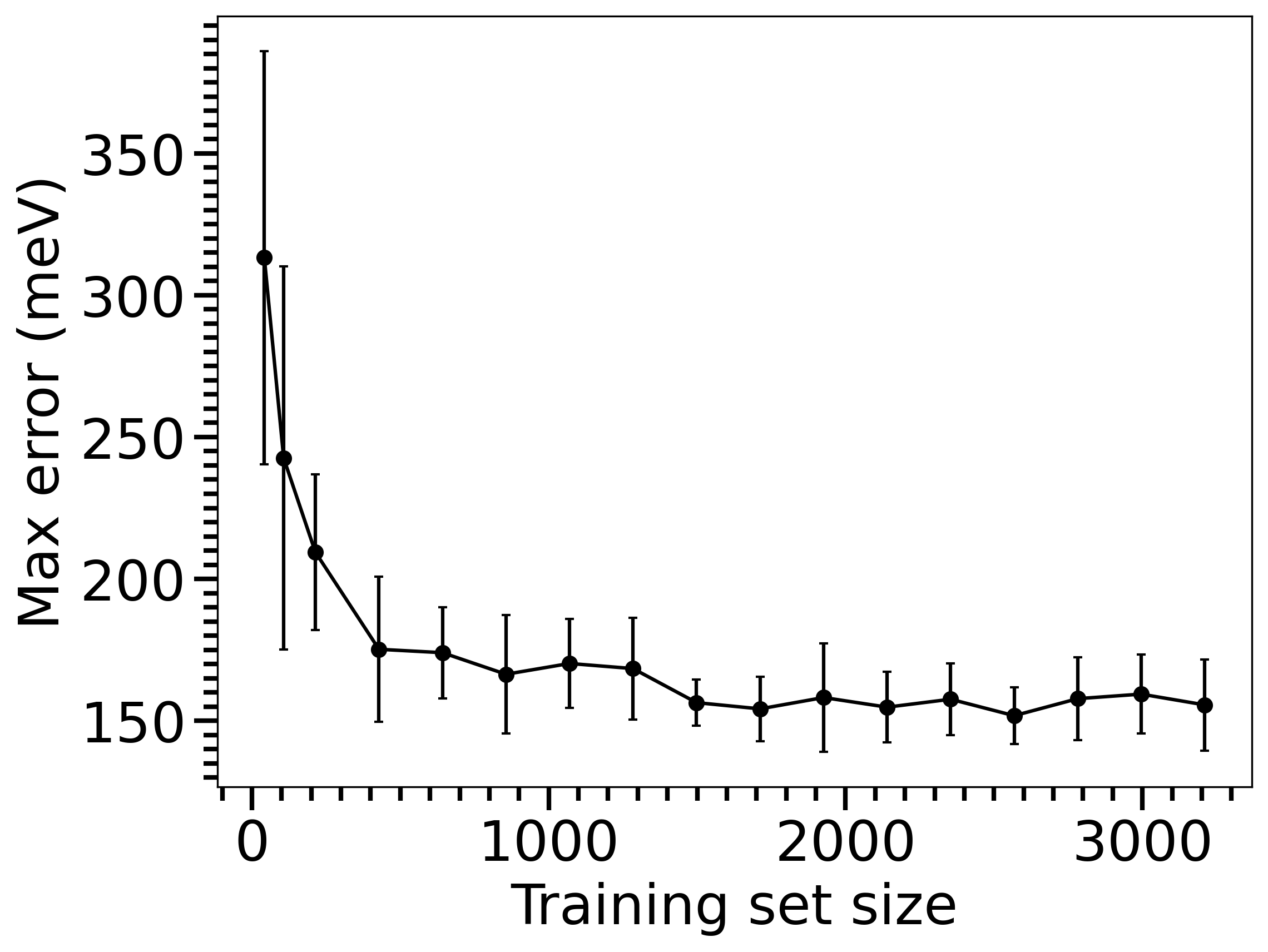}}
  \caption{\label{fig:figS3}Dependence of bandgap magnitude prediction (a) RMSE, (b) R$^2$, (c) MAE, and (d) Max error from SVR-RBF model on the number of training data. Error bars show standard deviations for 5 trials. The hyperparameters are optimized with RMSE metric.}
\end{figure}
\begin{figure}[!h]
\centering
  \subfloat[]{\label{fig:figS4a}\includegraphics[width=2.9in]{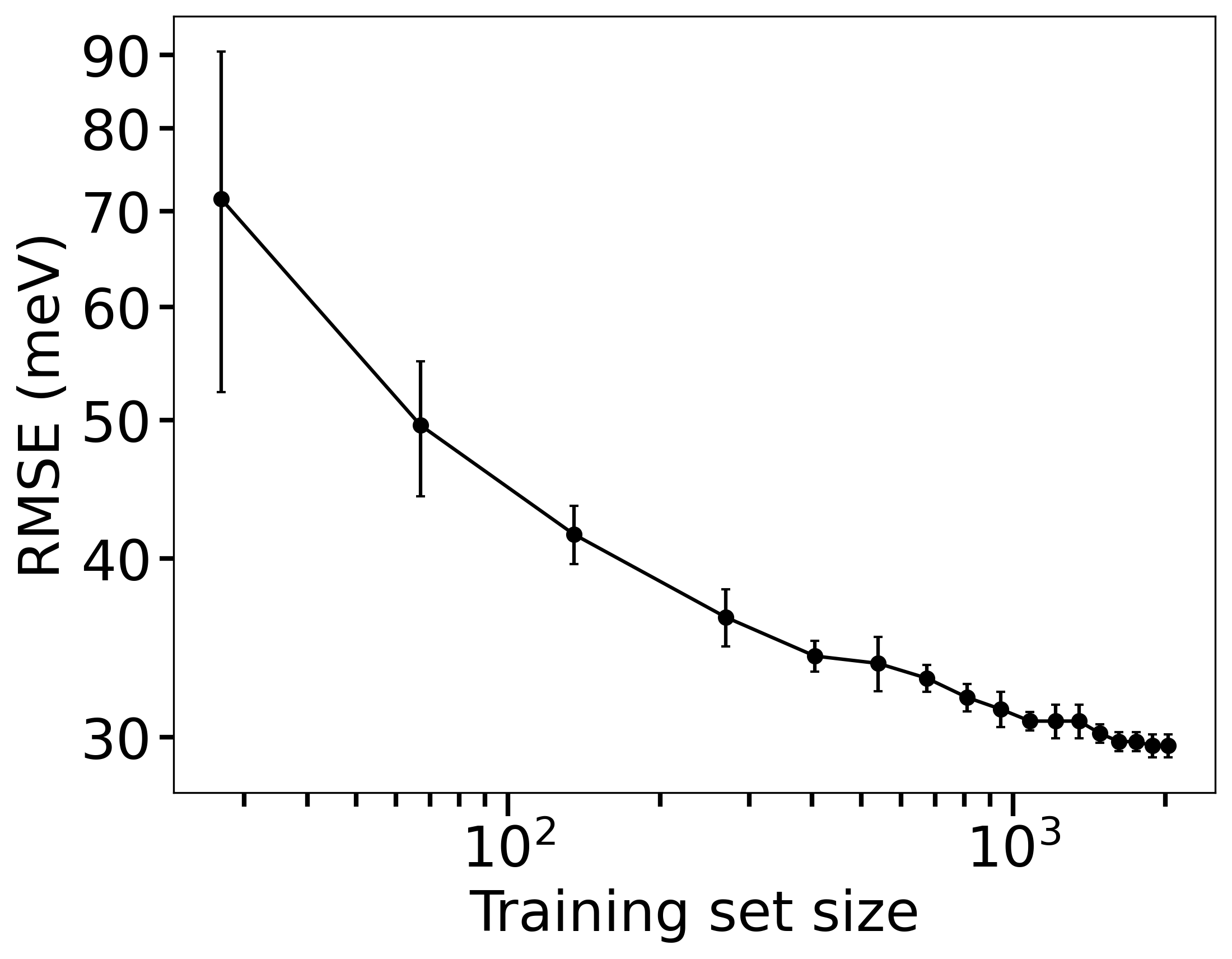}}
  \hspace{.3in}
  \subfloat[]{\label{fig:figS4b}\includegraphics[width=3.in]{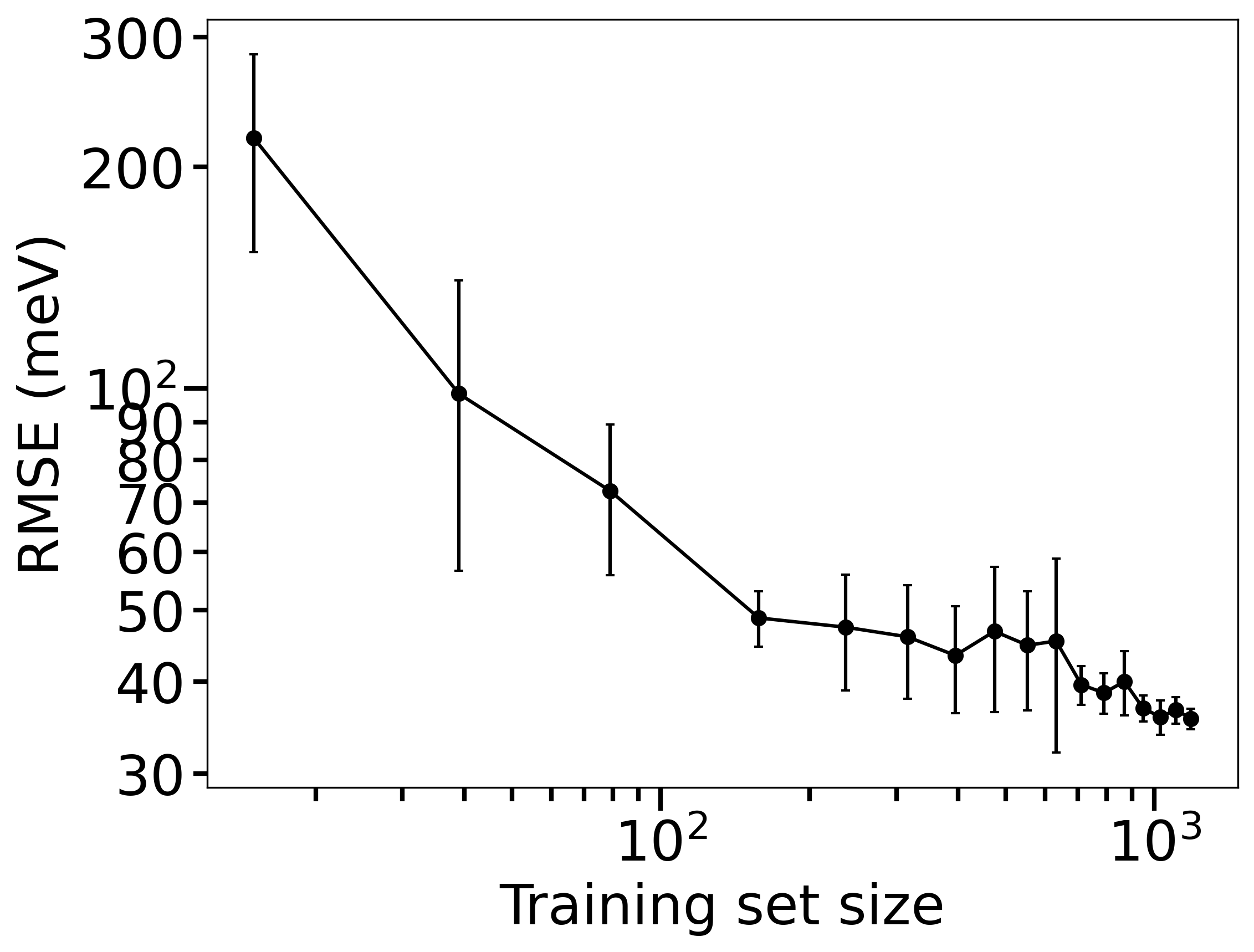}}\\
  \subfloat[]{\label{fig:figS4c}\includegraphics[width=3.1in]{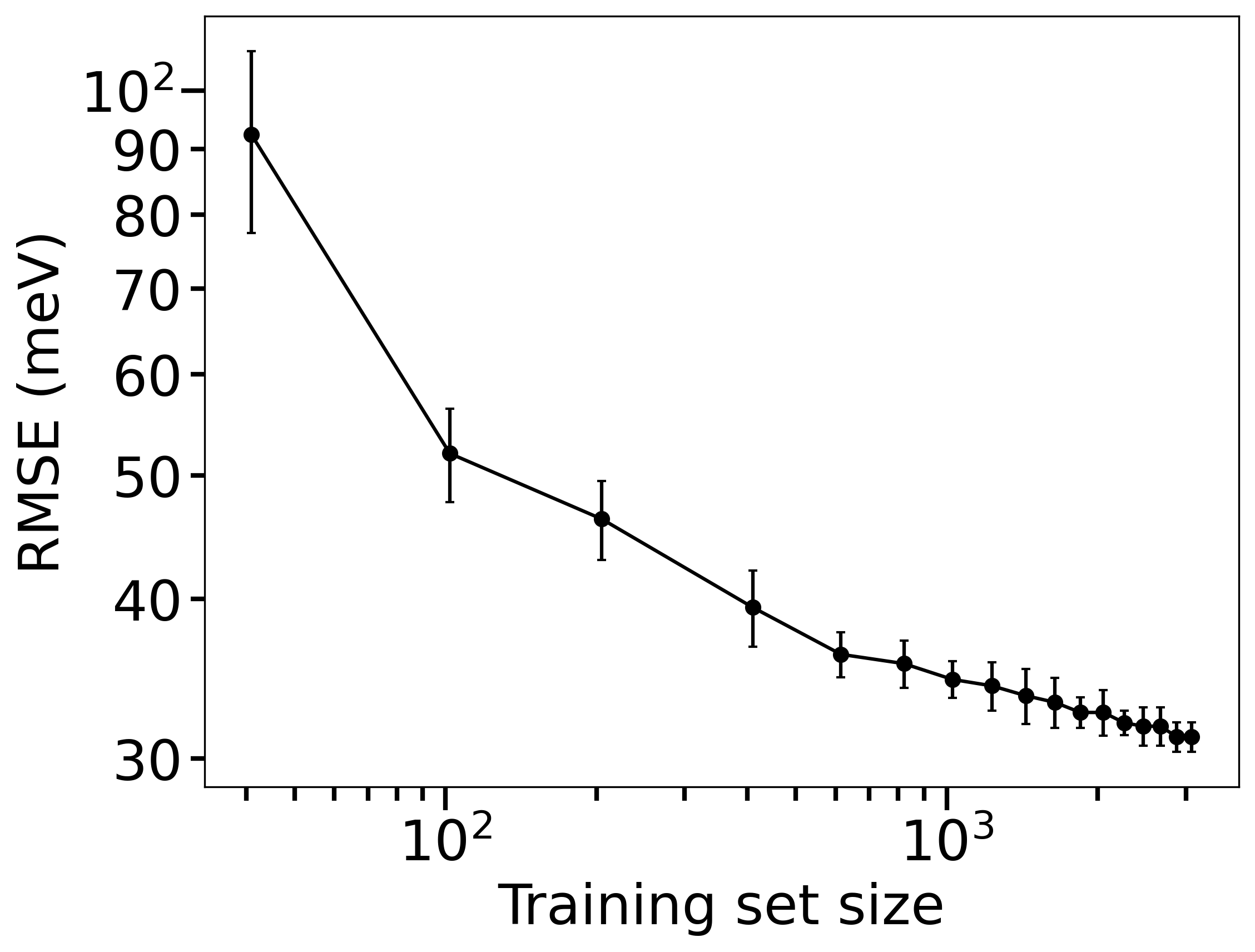}}
  \hspace{.3in}
  \subfloat[]{\label{fig:figS4d}\includegraphics[width=3.in]{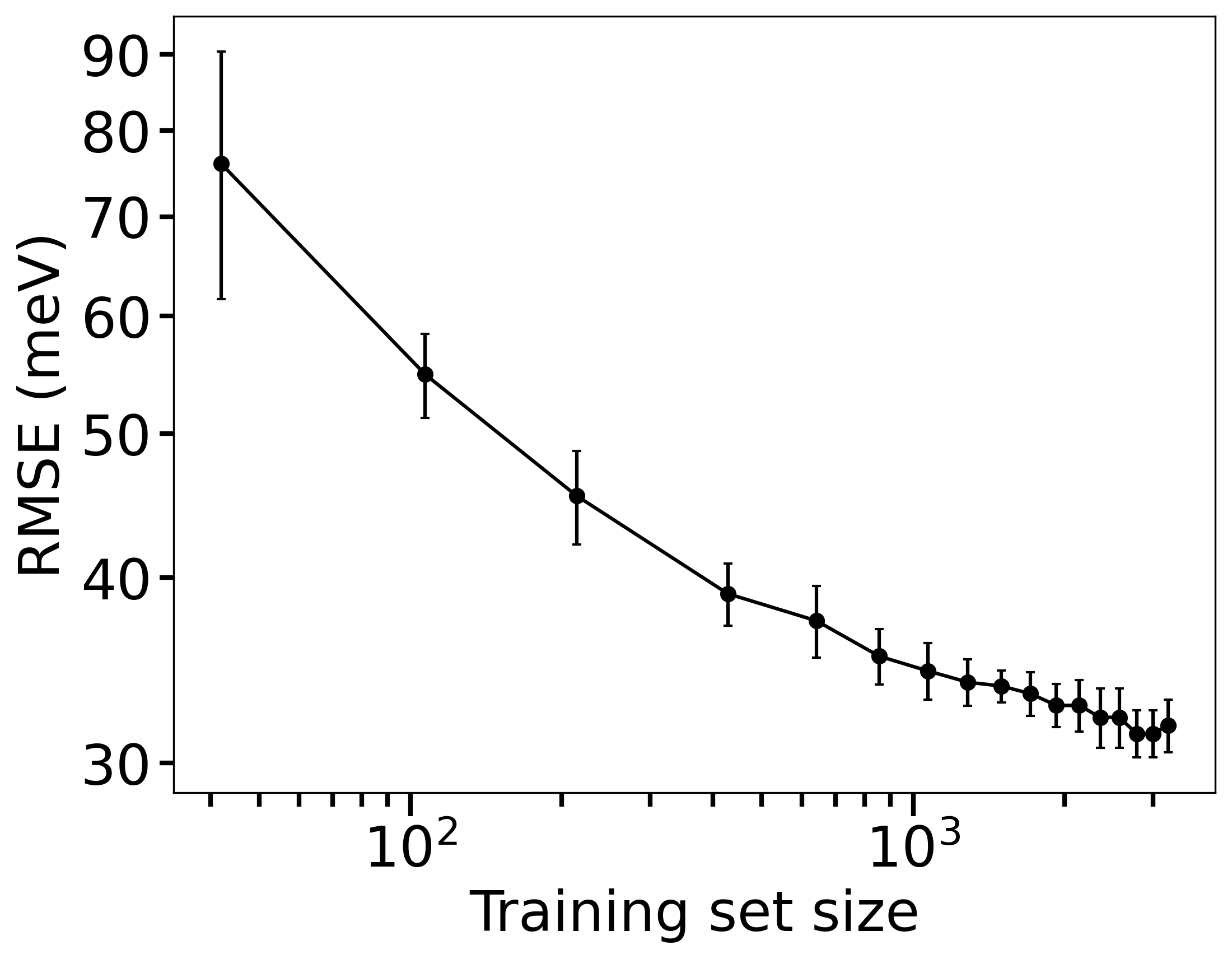}}
  \caption{\label{fig:figS4}Learning curve in log--log scale showing the performance of the SVR-RBF model for bandgap value predictions, with different dataset settings: (a) using the dataset that contains only direct bandgaps, (b) using the dataset that contains only indirect bandgaps, (c) using the dataset that contains both direct and indirect bandgaps while excluding the data points for which SVC-RBF model predicted an incorrect bandgap nature, and (d) using the original dataset. Figure (d) corresponds to Fig.~1a in the main manuscript. Error bars show standard deviations over 5 trials. The hyperparameters are optimized with RMSE metric.}
\end{figure}
\begin{figure}[!h]
    \centering
    \includegraphics[width=2.9in]{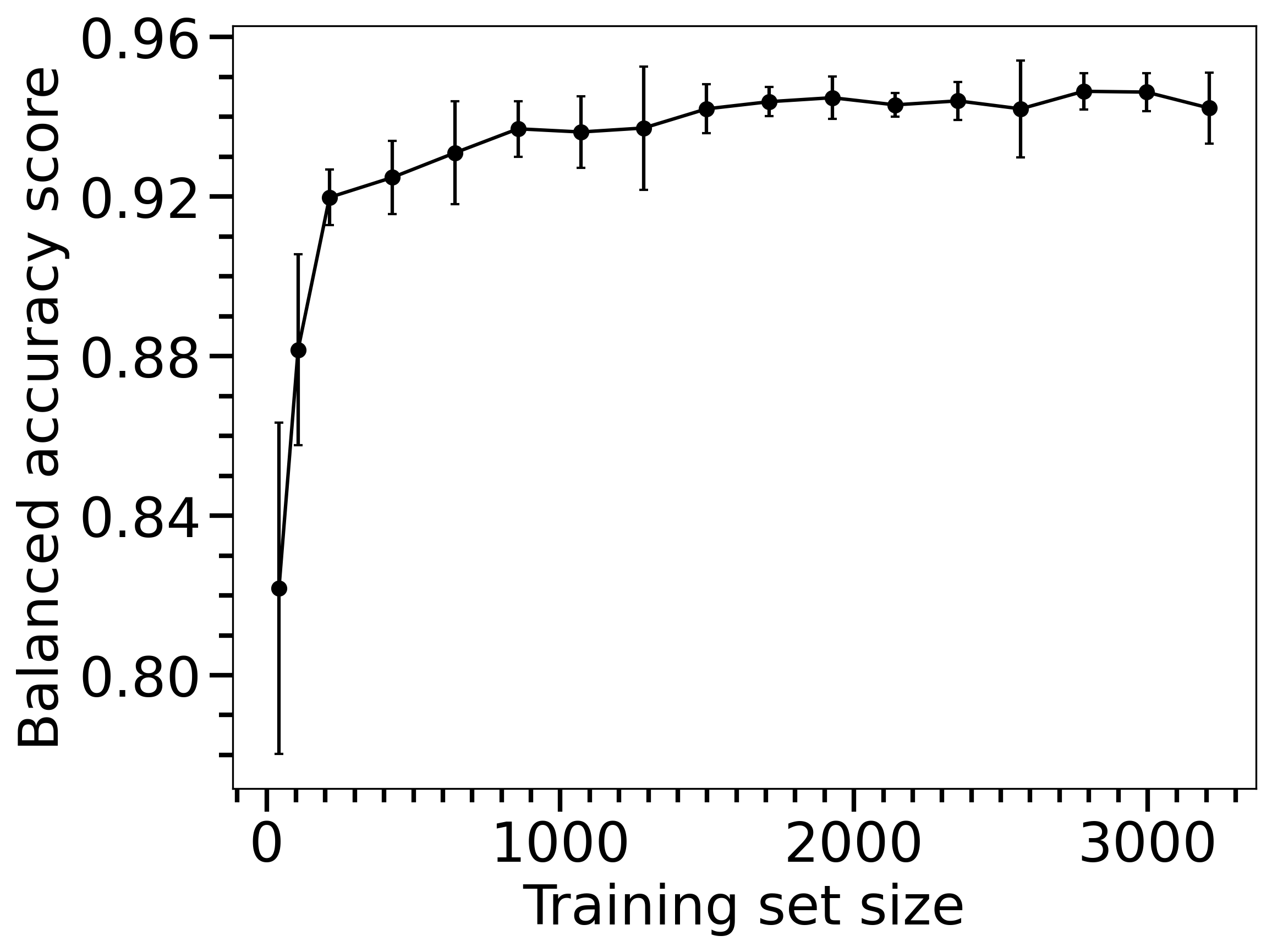}
    \caption{Dependence of bandgap nature prediction balanced-accuracy-score from SVC-RBF model on the number of training data. Error bars show standard deviations for 5 trials. Hyperparameters are optimized with accuracy-score metric.}
    \label{fig:figS5}
\end{figure}

\clearpage
\section{\label{sec:secS5}Bandgap prediction validations}

\begin{figure}[h!]
\centering
    \subfloat[]{\label{fig:figS6a}\includegraphics[height=3.in]{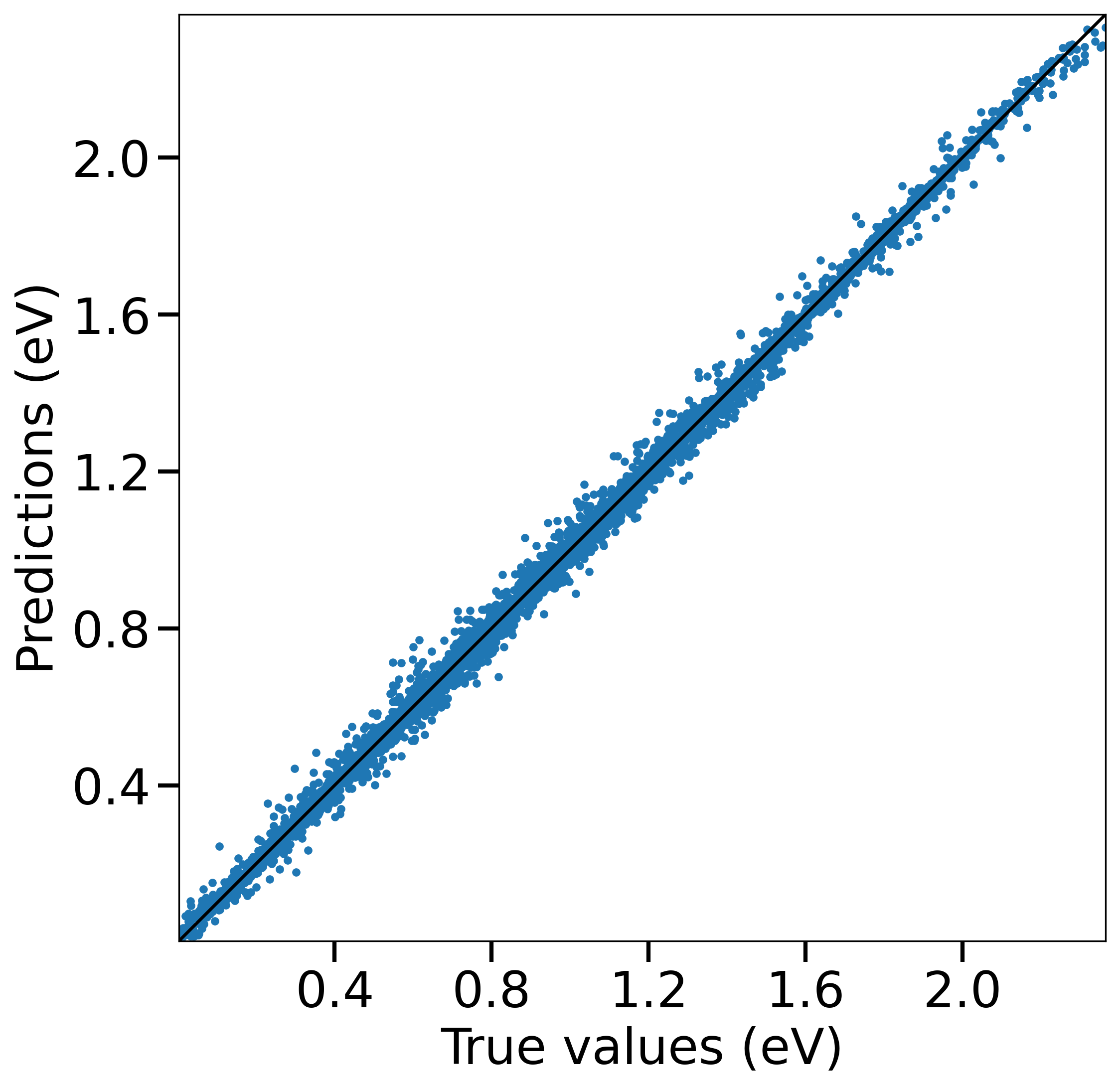}}
    \hspace{.2in}
    \subfloat[]{\label{fig:figS6b}\includegraphics[height=2.5in]{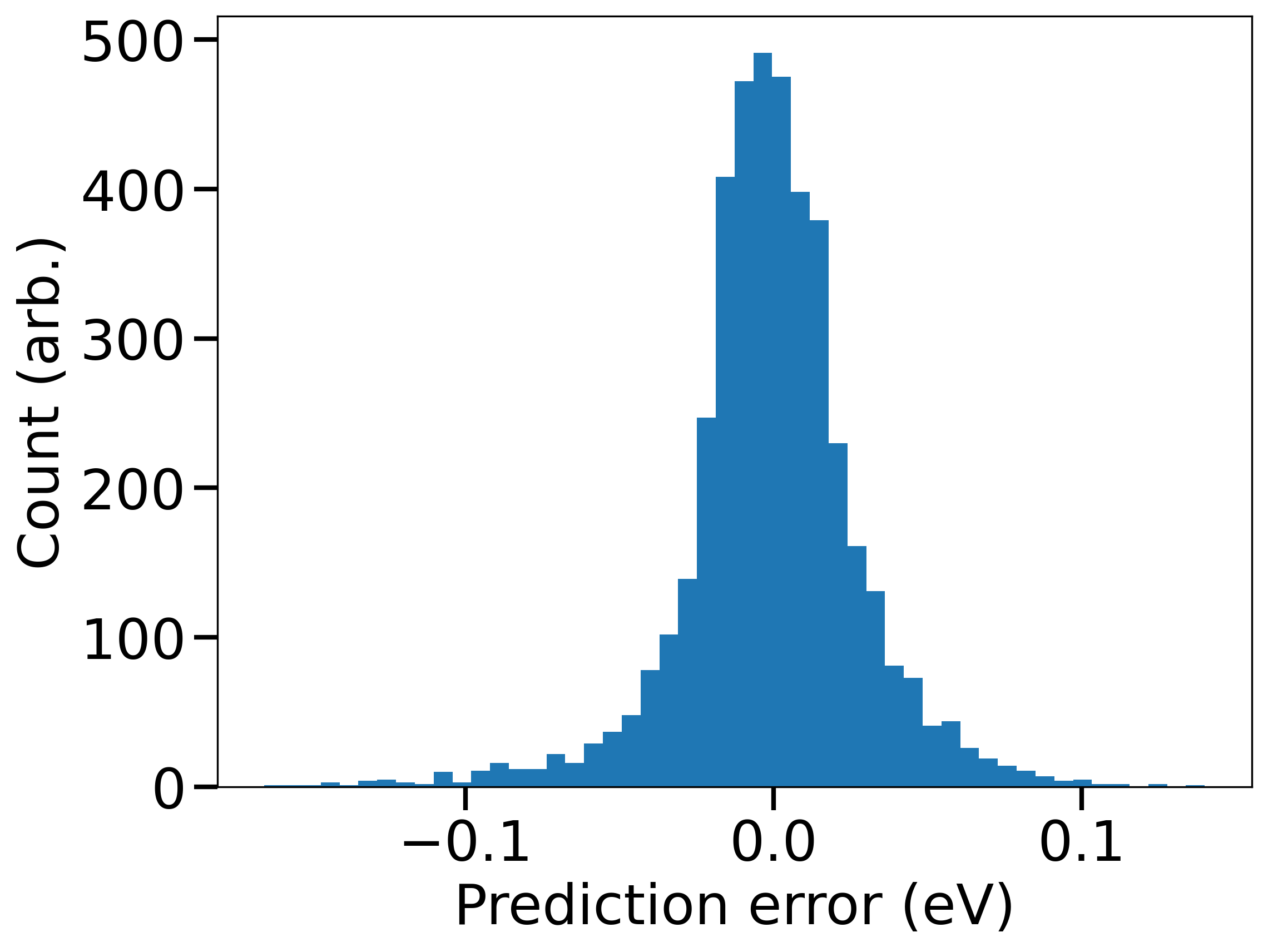}}
    \caption{\label{ig:figS6}The bandgap values prediction from the SVR-RBF models for all input data in the dataset. True values are the DFT calculated bandgap values. The predictions are the average bandgap values over the 5 model predictions from the trial set of the last point from Fig.~1a. (a) shows the comparison between true and predicted values, (b) is the prediction error (true value minus predicted value) distribution.}
\end{figure}

\begin{figure}[!h]
    \centering
    \includegraphics[width=3.in]{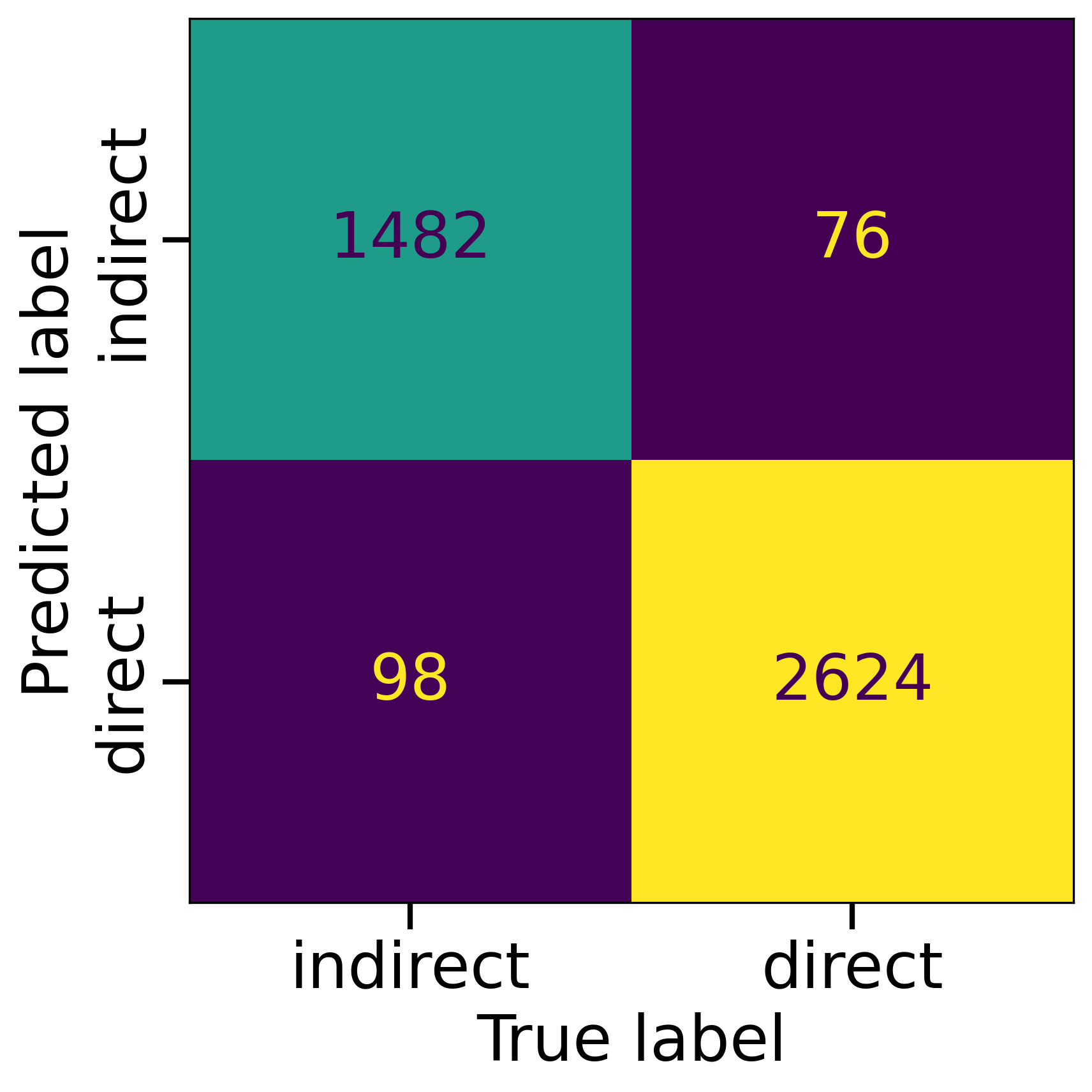}
    \caption{The confusion matrix of the bandgap nature predictions, showing the number of correct and wrong predictions for each class (direct and indirect). True labels are the bandgap nature determined from DFT calculations for all input data in the dataset. The prediction labels are the most frequent outcomes over the 5 SVC-RBF model predictions (mode value) from the trial set of the last point from Fig.~1b.}
    \label{fig:figS7}
\end{figure}

\clearpage

\section{\label{sec:secS6}Standard deviation distribution of bandgap prediction}

\begin{figure}[h!]
    \centering
    \includegraphics[width=3.in]{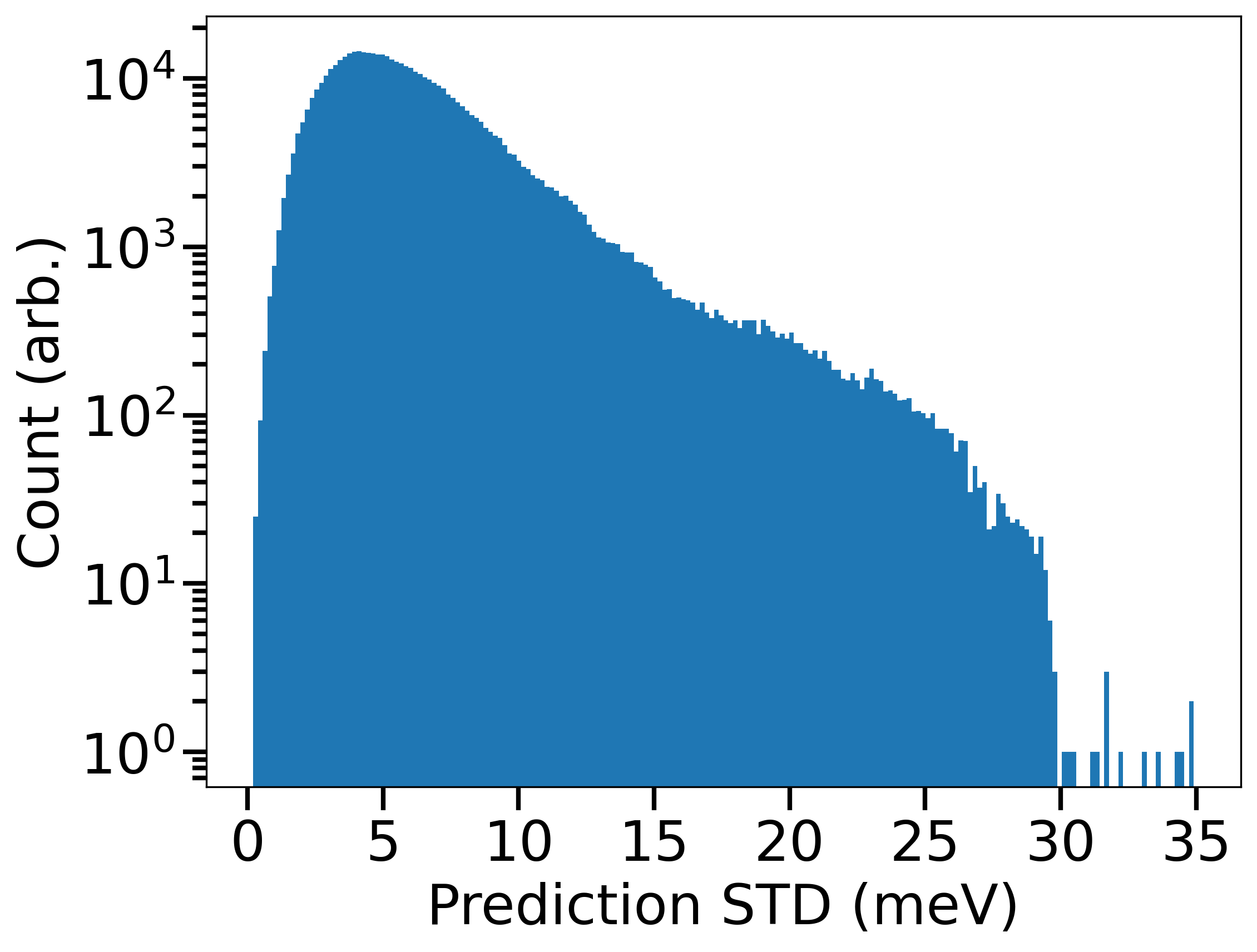}
    \caption{The standard deviation (STD) distribution of bandgap value predictions using the 5 SVR-RBF models from the last point of Fig.~1a. The figure is plotted in semi-log scale.}
    \label{fig:figS8}
\end{figure}

\section{\label{sec:secS7}Smoothening direct-indirect transition line}

\begin{figure}[h!]
    \centering
    \includegraphics[width=3.in]{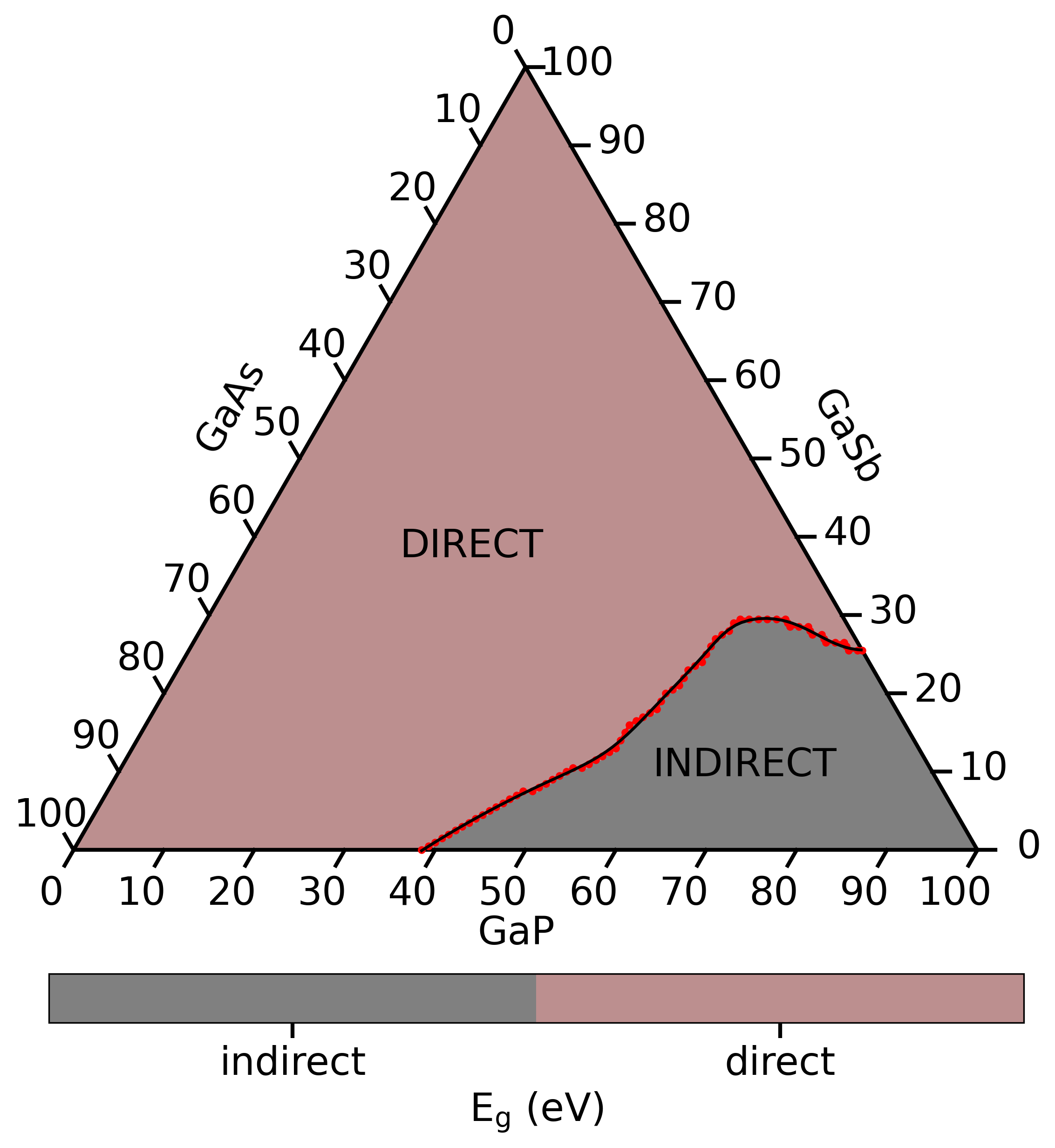}
    \caption{The mapping of bandgap nature for unstrained GaAsPSb (0.0\% strain). The labels `direct' and `indirect' describe the enclosed regions, with the nature of bandgap being direct and indirect, respectively. These areas are separated by the direct-indirect transition (DIT) points (red dots). The bandgap natures calculated are the most frequent outcomes over the 5 predictions (mode value) from the trial set of the last point from Fig.~1b. The calculated DIT points (red dots) are fitted (black line) with B-spline function with the smoothing factor, s=5.}
    \label{fig:figS9}
\end{figure}
We smoothen the calculated discrete direct-indirect transition (DIT) points with B-spline function \cite{Dierckx1982} (smoothing factor, s=5) as is implemented in scipy.interpolated \cite{2020SciPy-NMeth-short} routine.
\pagebreak

\begin{verbatim}
from scipy.interpolate import splprep, splev
x = DIT_x #x-coordinate of the calculated DITs
y = DIT_y #y-coordinate of the calculated DITs
tck, u = splprep([x, y], s=5)
smoothen_DIT_x, smoothen_DIT_y = splev(u, tck)
\end{verbatim}

\section{\label{sec:secS8}Bandgap values variation under strain for specific G\lowercase{a}A\lowercase{s}PS\lowercase{b}}

\begin{figure}[h!]
    \centering
    \includegraphics[width=3.in]{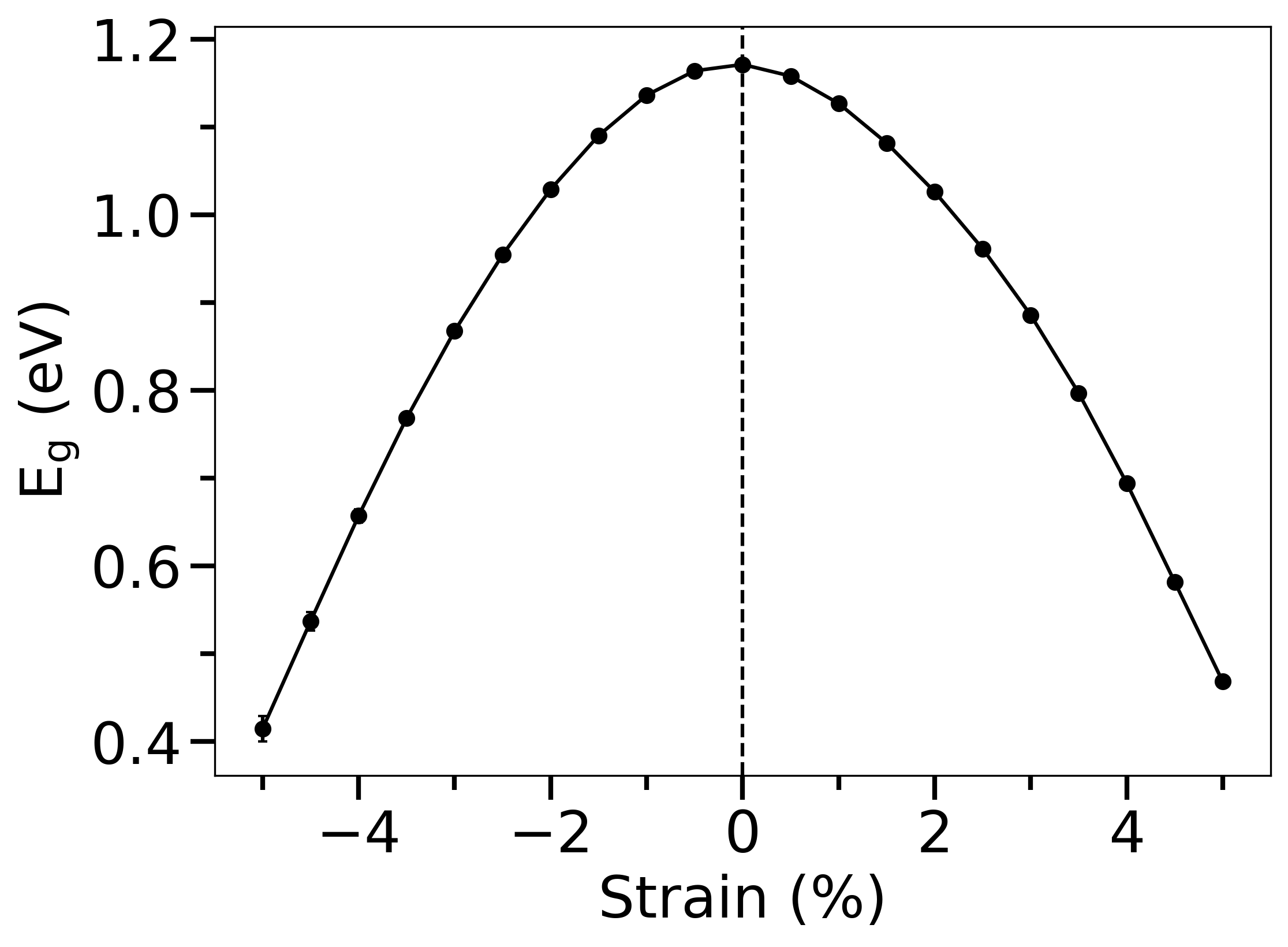}
    \caption{The variation of bandgap values (E$_{\text{g}}$) under biaxial strain in GaAs$_{0.333}$P$_{0.333}$Sb$_{0.334}$. The positive and negative strain values indicate the tensile and compressive strains, respectively. The bandgap values are the average values over the 5 model predictions from the trial set of the last point from Fig.~1a and error bars show standard deviations.}
    \label{fig:figS10}
\end{figure}

\clearpage
\section{\label{sec:secS9}Substrate effect in G\lowercase{a}A\lowercase{s}PS\lowercase{b} bandgap phase diagram}

\begin{figure}[h!]
    \centering
    \subfloat[GaP substrate]{
    \includegraphics[width=3.in]{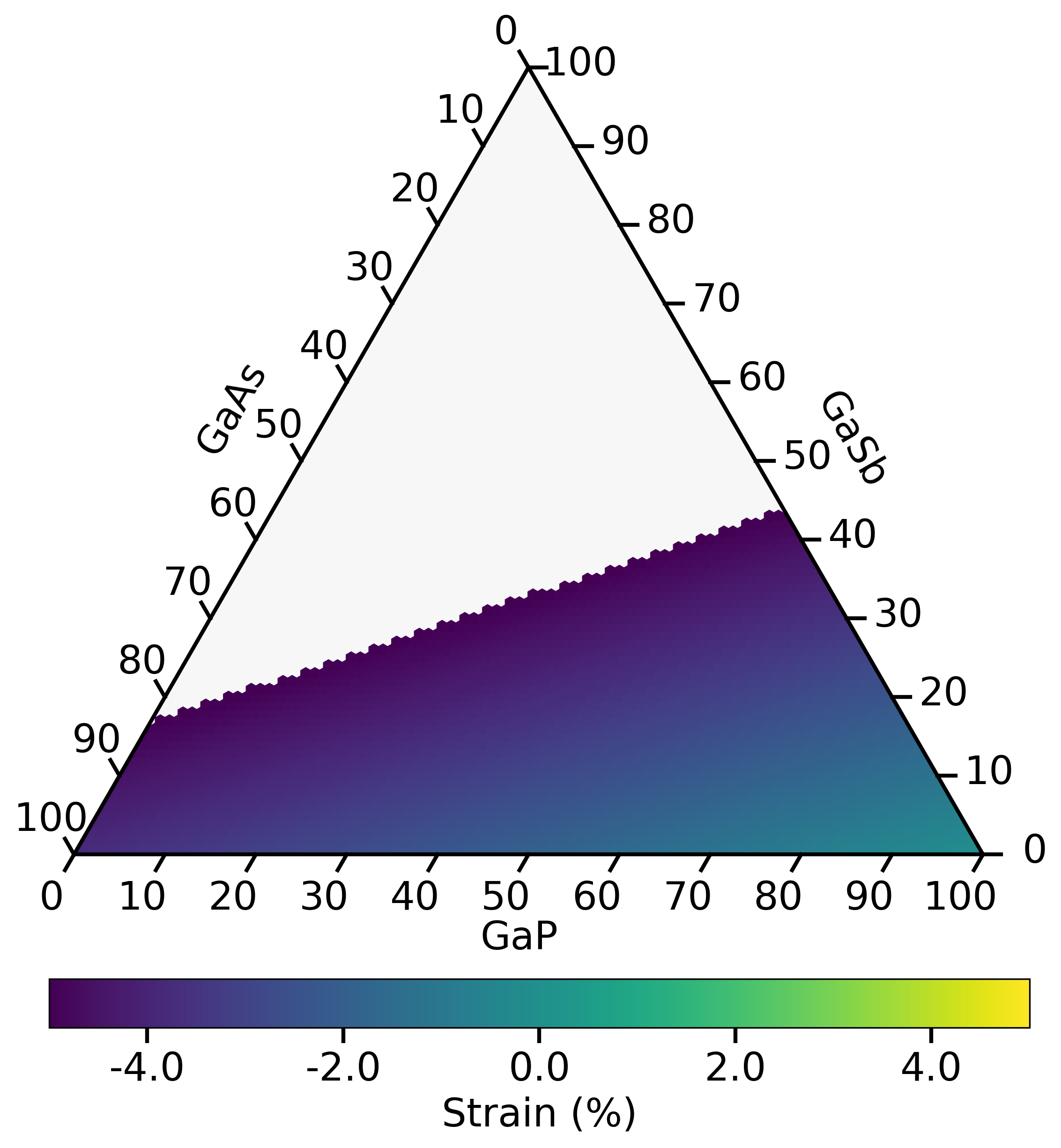}
	\hspace{.4in}
	\includegraphics[width=3.in]{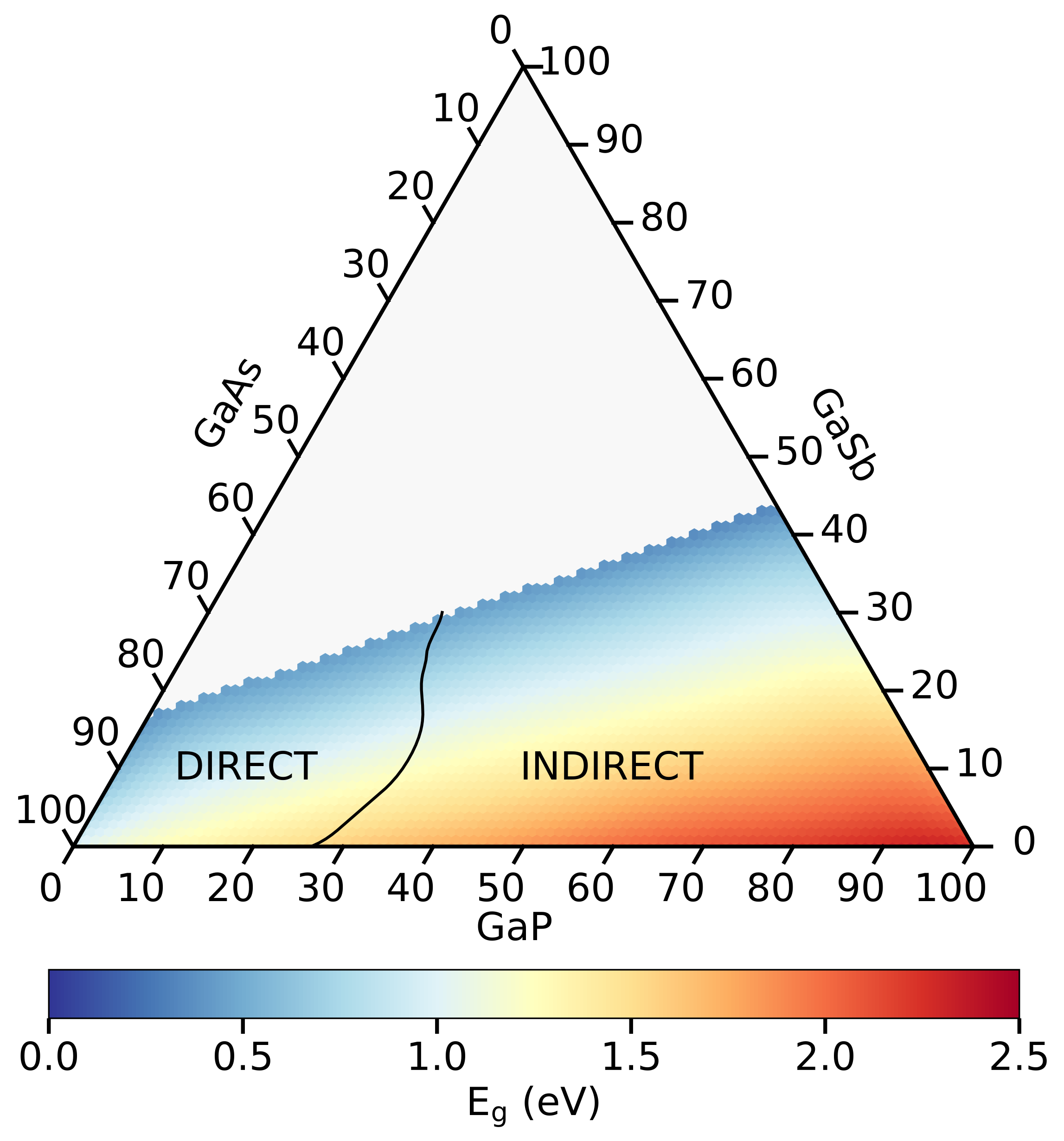}}\\
	\subfloat[GaSb substrate]{
    \includegraphics[width=3.in]{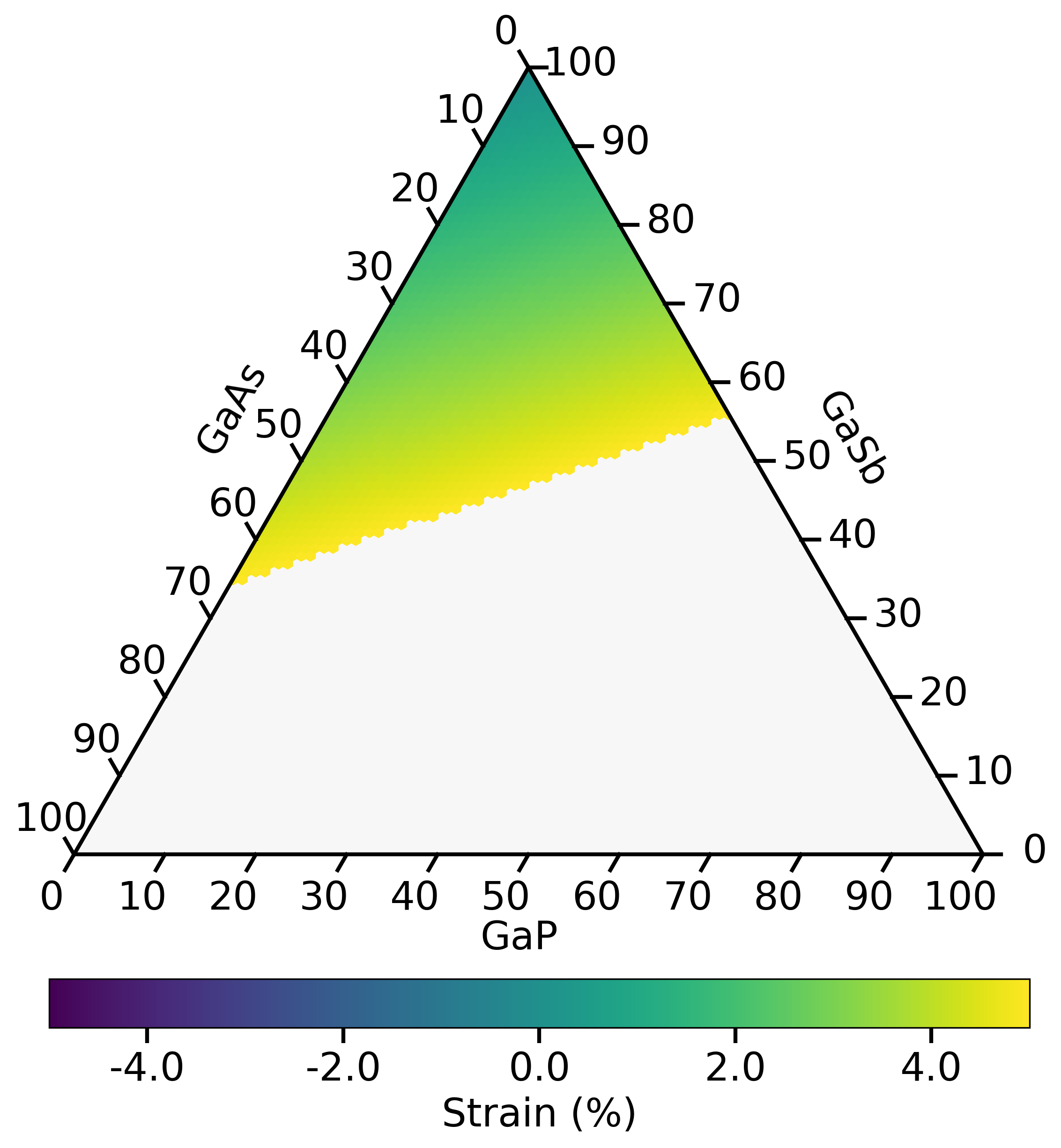}
	\hspace{.4in}
	\includegraphics[width=3.in]{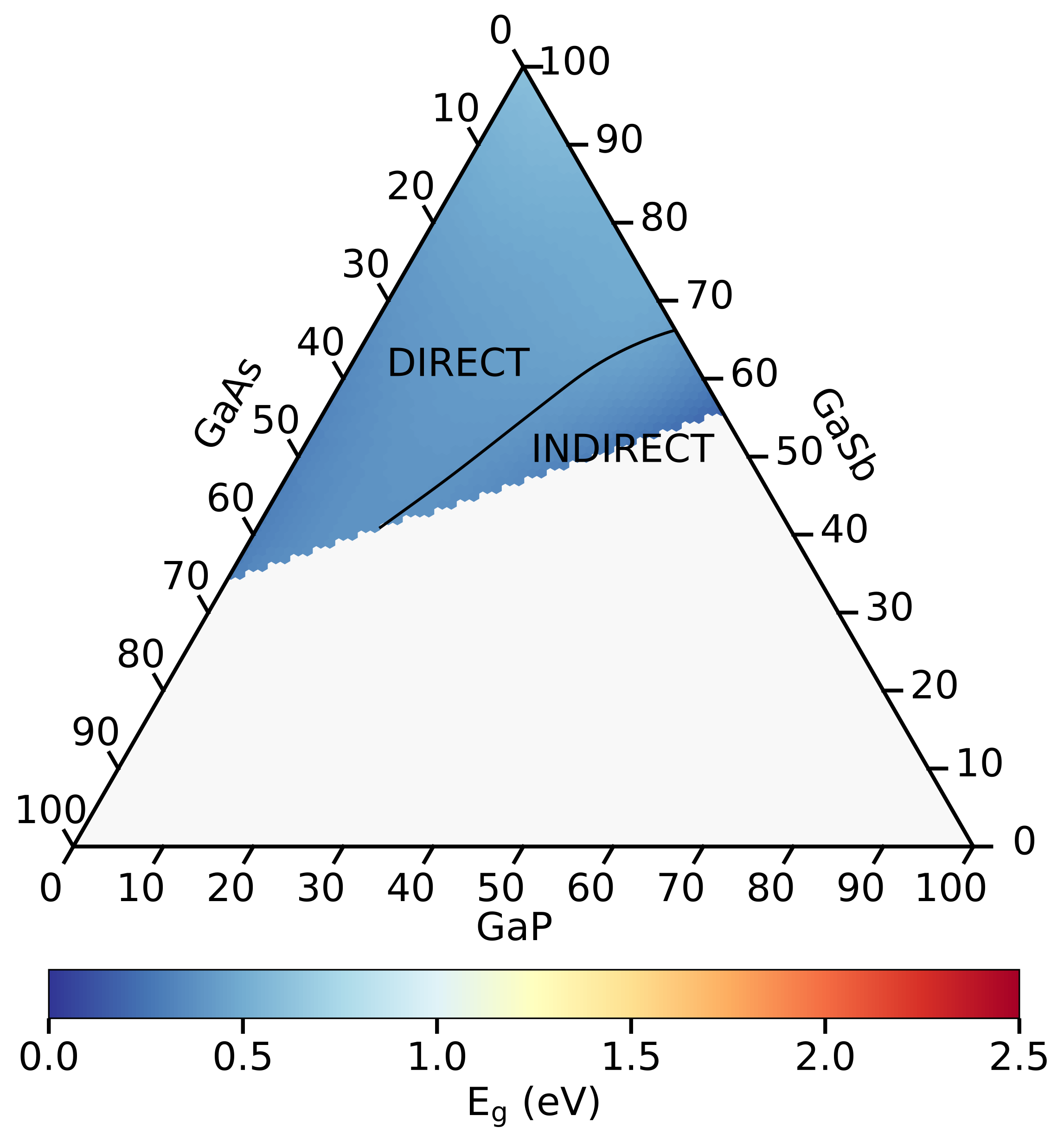}}
    \caption{The effect of substrate on GaAsPSb epi-layer under the `theoretical epitaxy' model \cite{Mondal2022,Mondal2023} (up to $5\%$ compressive and tensile strain). The left column shows the calculated biaxial strain values using Eqs.~8 and 9. The right column presents predicted bandgap values in color (E$_{\text{g}}$). The labels `direct' and `indirect' describe the enclosed regions, with the nature of bandgap being direct and indirect, respectively. The bandgap values are the average values over the 5 model predictions from the trial set of the last point from Fig.~1a. The nature of the bandgaps are the most frequent outcomes over the 5 predictions (mode value) from the trial set of the last point from Fig.~1b.}
\end{figure}
\begin{figure}[h!]\ContinuedFloat
    \centering
    \subfloat[InP substrate]{
    \includegraphics[width=3.in]{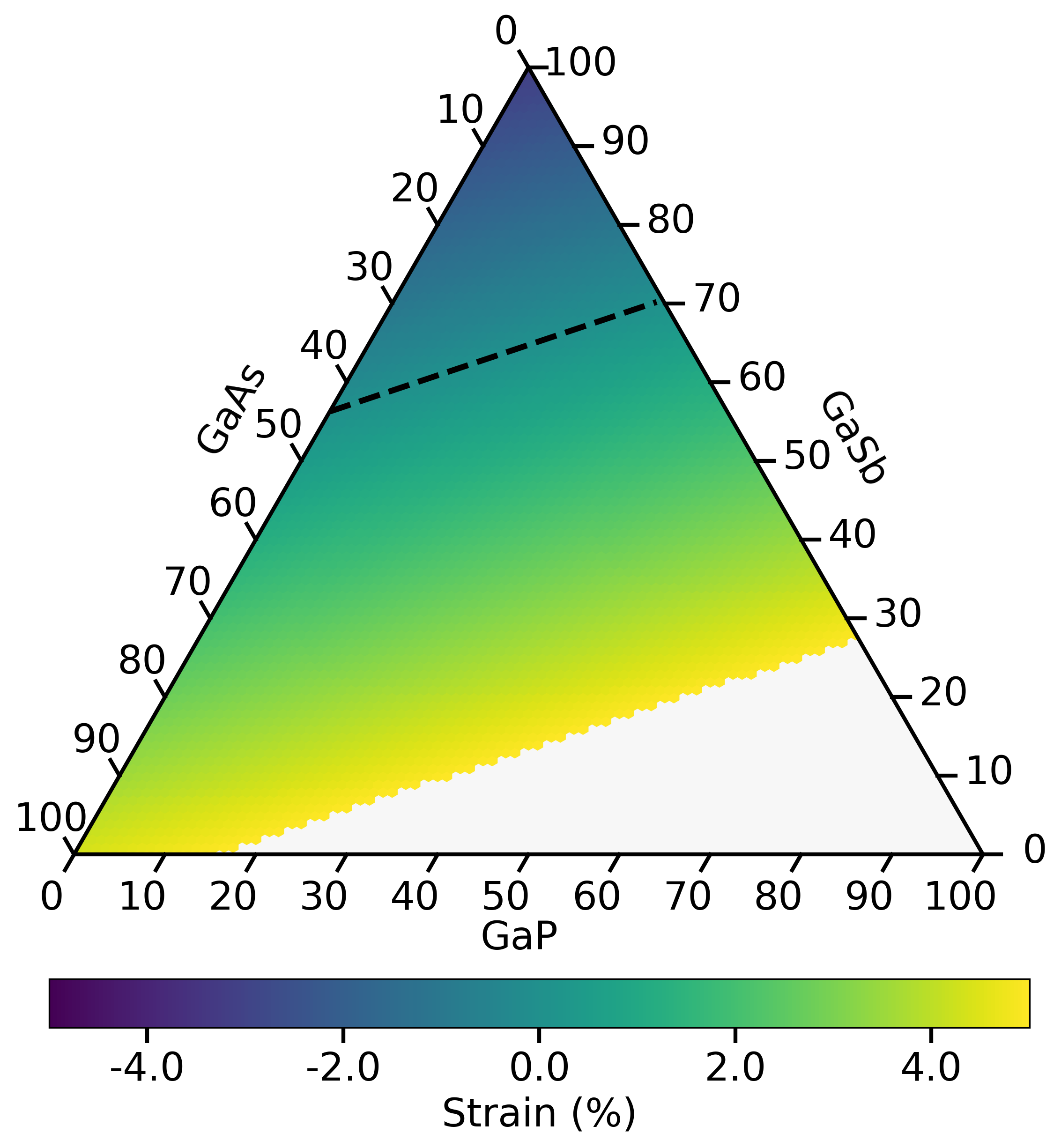}
	\hspace{.4in}
	\includegraphics[width=3.in]{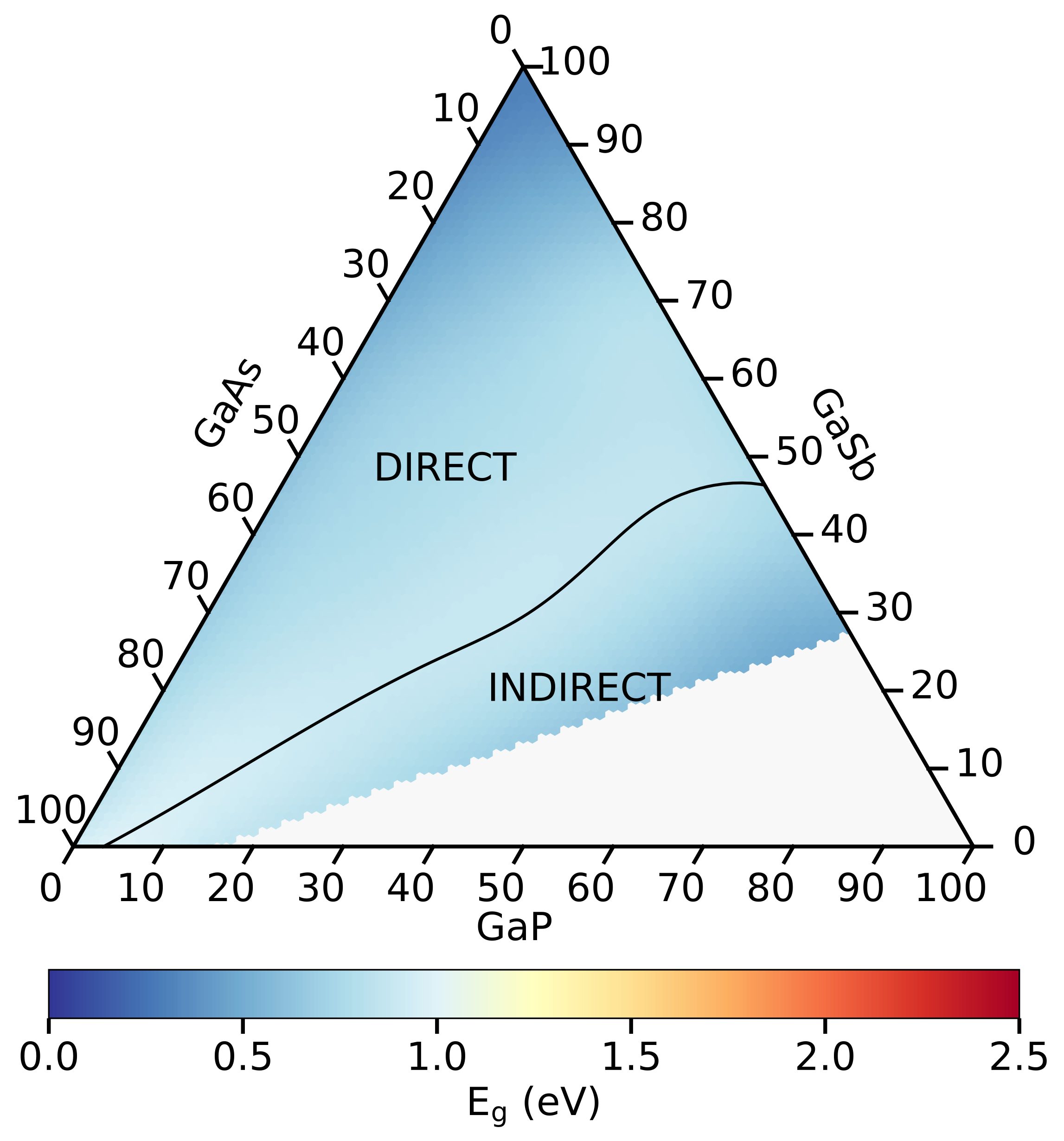}}\\
	\subfloat[Si substrate]{
    \includegraphics[width=3.in]{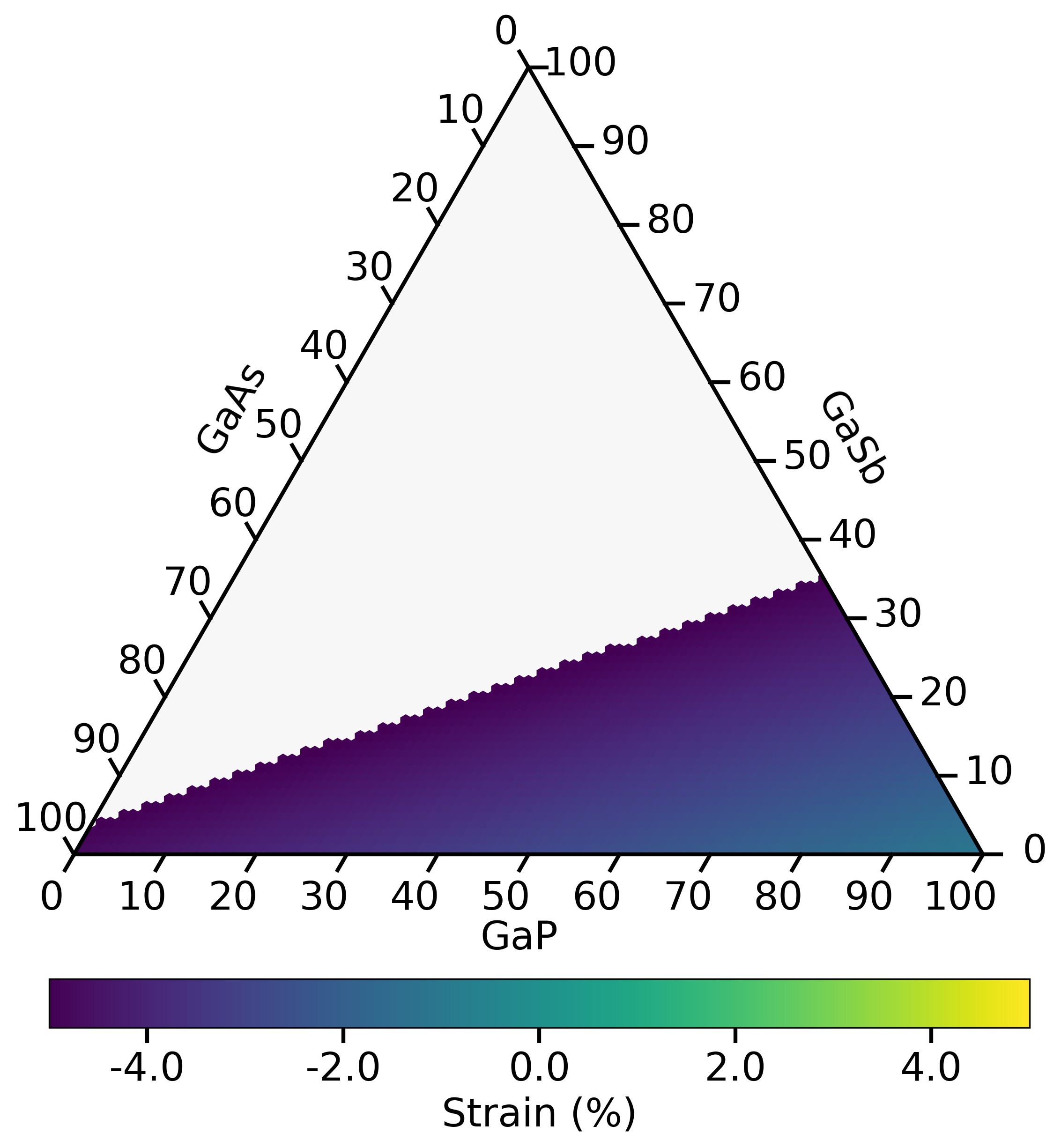}
	\hspace{.4in}
	\includegraphics[width=3.in]{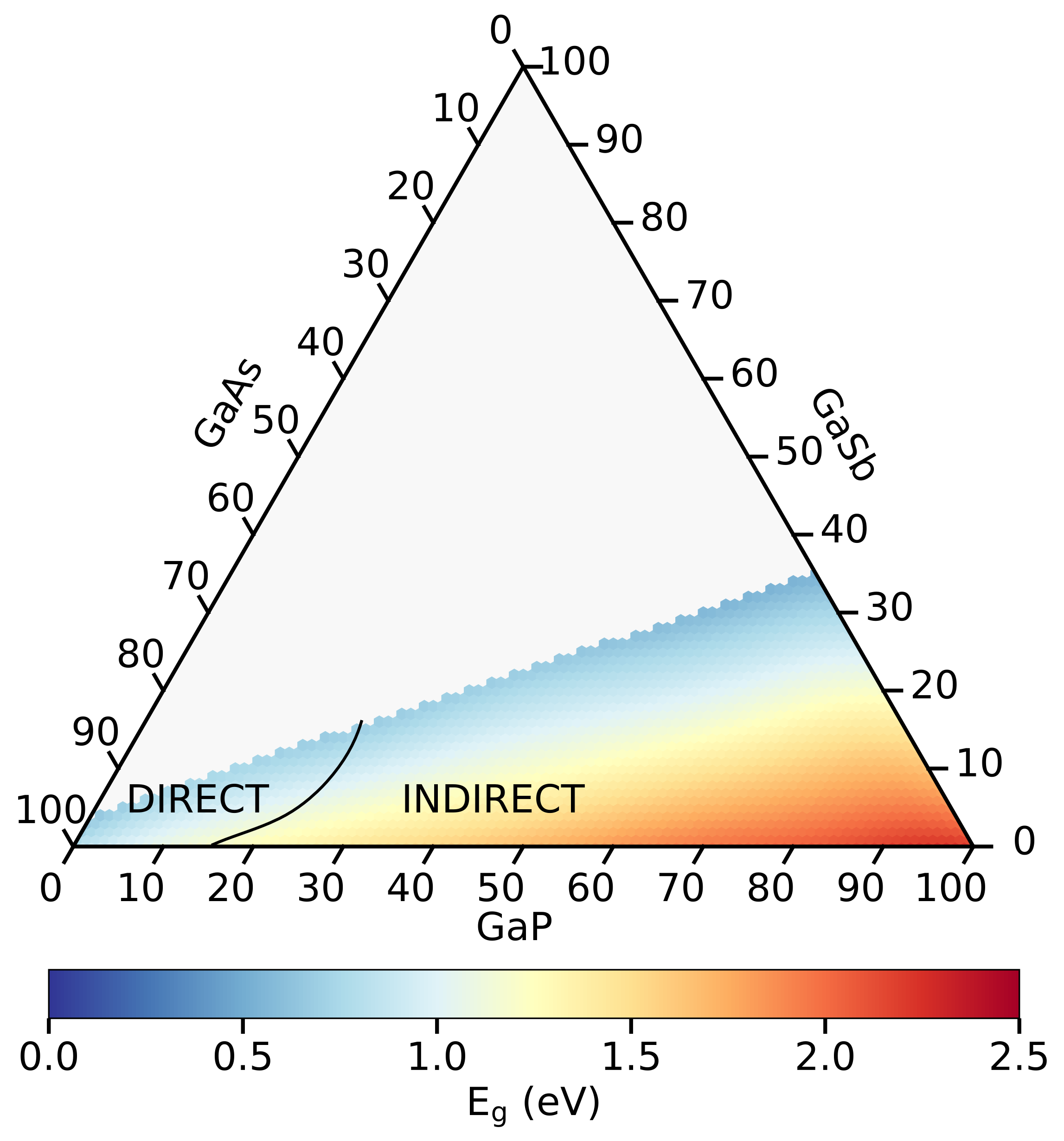}}
    \caption{\textit{(continue)} The effect of substrate on GaAsPSb epi-layer under the `theoretical epitaxy' model \cite{Mondal2022,Mondal2023} (up to $5\%$ compressive and tensile strain). The left column shows the calculated biaxial strain values using Eqs.~8 and 9. The black dotted line in (c) indicates the perfect lattice matching (no strain) compositions. The right column presents predicted bandgap values in color (E$_{\text{g}}$). The labels `direct' and `indirect' describe the enclosed regions, with the nature of bandgap being direct and indirect, respectively. The bandgap values are the average values over the 5 model predictions from the trial set of the last point from Fig.~1a. The nature of the bandgaps are the most frequent outcomes over the 5 predictions (mode value) from the trial set of the last point from Fig.~1b.}
    \label{fig:figS11}
\end{figure}
\clearpage
\bibliography{supplement}